\documentclass[alpha-refs]{wiley-article} %num-refs

\usepackage{graphicx}
\usepackage{lmodern}
\usepackage[space]{grffile}
\usepackage{latexsym}
\usepackage{textcomp}
\usepackage{longtable}
\usepackage{tabulary}
\usepackage{booktabs,array,multirow}
\usepackage{amsfonts,amsmath,amssymb}
\usepackage{natbib}
\usepackage{url}
\usepackage{hyperref}
\hypersetup{colorlinks=false,pdfborder={0 0 0}}
\usepackage{etoolbox}

\usepackage{xcolor}

\newif\iflatexml\latexmlfalse

\AtBeginDocument{\DeclareGraphicsExtensions{.pdf,.PDF,.eps,.EPS,.png,.PNG,.tif,.TIF,.jpg,.JPG,.jpeg,.JPEG}}

\usepackage[utf8]{inputenc}
\usepackage[english]{babel}

\usepackage{amsmath} %ESF
\usepackage{longtable} %ESF
\usepackage{lscape} %ESF
\usepackage{pdflscape} %ESF

\iflatexml
\else
\paperfield{}
\corraddress{School of Mathematical Sciences. Queensland University of Technology. Australia}
\corremail{santosfe@qut.edu.au; edgar.santosfdez@gmail.com;
k.mengersen@qut.edu.au}
\fundinginfo{This research was supported by the Australian Research Council (ARC) Laureate Fellowship Program under the project ``Bayesian Learning for Decision Making in the Big Data Era'' (ID: FL150100150)}
\fi

\papertype{Original Article}

\title{
Increasing trust in new data sources: crowdsourcing image classification for ecology.
}

\author[1,2]{Edgar Santos-Fernandez }  %\emailx{santosfe@qut.edu.au; edgar.santosfdez@gmail.com}
\affil[1]{School of Mathematical Sciences. Queensland University of Technology. Australia.}

\affil[2]{Centre for Data Science. Queensland University of Technology. Australia.}

\author[1,2]{Julie Vercelloni}

\author[1,2]{Aiden Price}  

\author[1,2]{Grace Heron} 

\author[3]{Bryce Christensen}  

\affil[3]{Visualisation and Interactive Solutions for Engagement and Research (VISER) Lab. Queensland University of Technology.
Australia.}

\author[1,2]{Erin E. Peterson}  

\author[1,2]{Kerrie Mengersen}

\runningauthor{Edgar Santos-Fernandez et al.}

\begin{document}

\maketitle
\selectlanguage{english}

\begin{abstract}

Crowdsourcing methods facilitate the production of scientific information by non-experts. This form of citizen science (CS) is becoming a key source of complementary data in many fields to inform data-driven decisions and study challenging problems. However, concerns about the validity of these data often constrain their utility. 
In this paper, we focus on the use of citizen science data in addressing complex challenges in environmental conservation. 
We consider this issue from three perspectives. 
First, we present a literature scan of papers that have employed Bayesian models with citizen science in ecology. Second, we compare several popular majority vote algorithms and introduce a Bayesian item response model that estimates and accounts for participants' abilities after adjusting for the difficulty of the images they have classified. The model also enables participants to be clustered into groups based on ability. 
Third, we apply the model in a case study involving the classification of corals from underwater images from the Great Barrier Reef, Australia.
We show that the model achieved superior results in general and, for difficult tasks, a weighted consensus method that uses only groups of experts and experienced participants produced better performance measures.
Moreover, we found that participants learn as they have more classification opportunities, which substantially increases their abilities over time.
Overall, the paper demonstrates the feasibility of CS for answering complex and challenging ecological questions when these data are appropriately analysed. 
This serves as motivation for future work to increase the efficacy and trustworthiness of this emerging source of data.
\end{abstract}%

\section{Introduction \label{section intro}}

\begin{figure}[ht]
  \centering
  \includegraphics[width=3.5in]{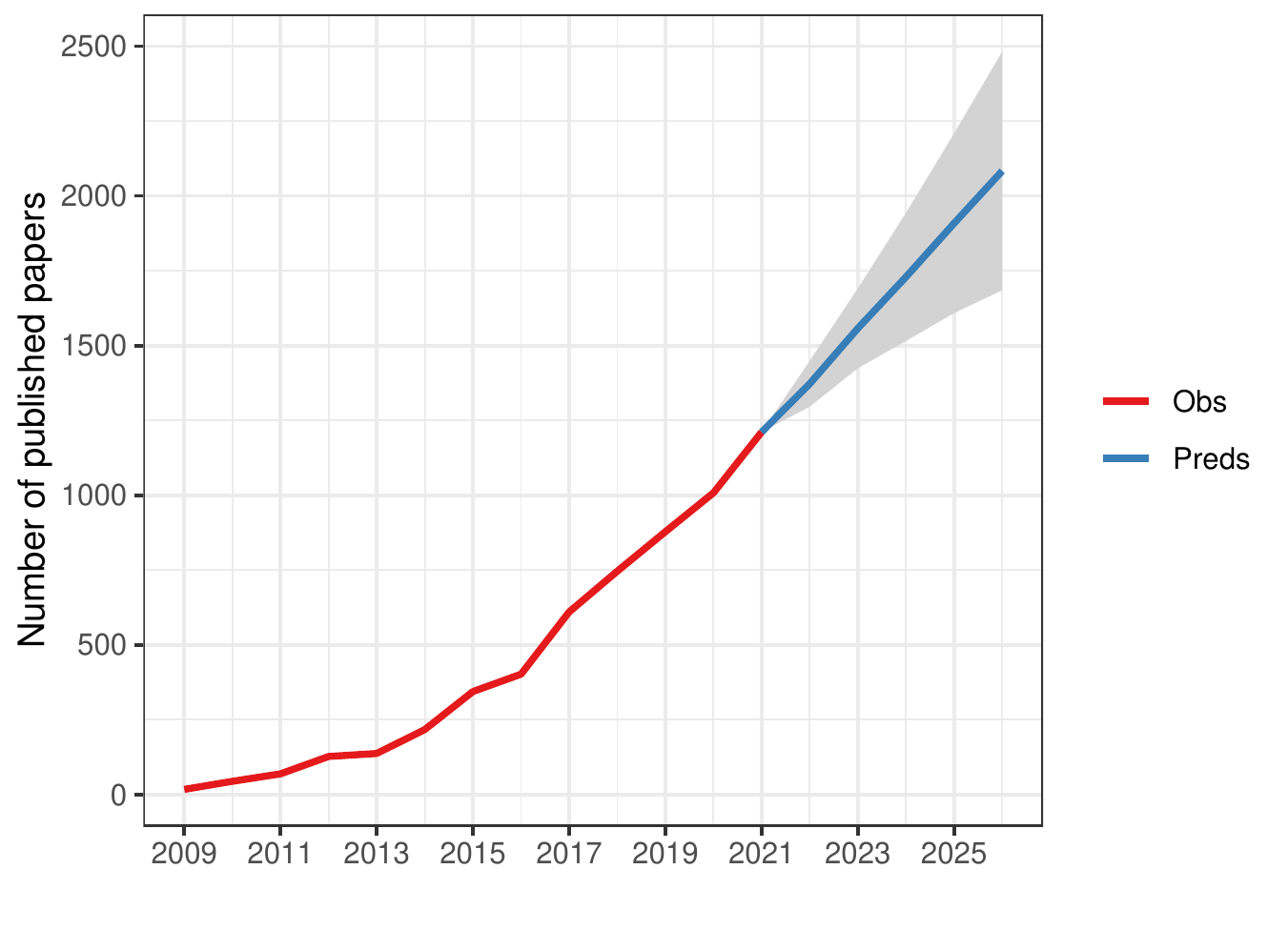}
\caption{Number of citizen science published articles per year since 2009 according to the Web of Science database and the forecast for the next five years with confidence intervals.}
\label{fig:papers}
\end{figure}

Citizen science (CS) and crowdsourcing involve volunteers and amateur scientists in the scientific process.
Over the last decade the number of publications using CS data has increased dramatically (Fig~\ref{fig:papers}).
CS projects are now entrenched in many fields of science,  including the notification and classification of astronomical events to learn about our universe, the identification and monitoring of environmental phenomena to learn about our world, and the reporting and assessment of health and medical outcomes to learn about our own selves.  
Quantitative information obtained from citizen scientists can be analysed in its own right, or it can fill gaps in professional data collection programs or experimental studies, or it can be used for complementary purposes such as training machine learning algorithms
\citep{bradter2018can}.

In the field of ecology, millions of citizen scientists are engaged in research projects worldwide, producing volumes of information in a cost-effective and timely manner and contributing substantively to aspirational targets such as the United Nations Sustainable Development Goals \citep{hsu2014development, fritz2019citizen}.  These projects embrace a wide range of challenges, including assessments of climate change impacts, water and air quality measurements, monitoring of biodiversity trends and patterns in abundance,  distribution, and richness of native and exotic species.
Citizen science platforms have also become established. For example, iNaturalist, eButterfly and Zooniverse host hundreds of projects involving image classification.
These online platforms have gained substantial recognition in recent years because they have the potential to reduce the workload of ecological experts and they engage large communities of contributors.

Notwithstanding the popularity and benefits of CS, considerable data quality concerns remain regarding research involving data elicited by non-expert citizen scientists \citep{downs2021perspectives}.
These concerns primarily focus on the potential for bias arising from the unstructured nature of the data and the differing abilities of the participants.
For example, while the individual classification accuracy is high in some CS image classification projects, ranging from 70\% to 95\%  \citep[e.g.][]{kosmala2016assessing}, aggregation via some form of consensus is often required to produce reliable classifications and estimates \citep{santos2020correcting, santos2021}.
Many aggregation methods, including majority voting, enjoy great success in the literature.  
These methods rely on the assumption that participants have greater than 50\% chance of answering correctly.
However, such an assumption is not always satisfied in the presence of difficult tasks, in which the majority of the classifications can be incorrect \citep{raykar2010learning}.

In this paper, we focus on a particular set of problems that are of great interest for citizen science in ecology, namely the classification of objects in photos and videos. 
We investigate the feasibility of crowdsourcing methods for citizen science as a viable solution for manual classification of images when the task is difficult. This may occur, for example, if the images represent complex ecosystems, depicting aggregations of diverse species communities that change in space and time. Difficult tasks can also be related to images produced by different sources, such as environmental monitoring programs and cameras.  

We consider this problem through a Bayesian lens. To set the work in context, we first present a scan of papers that have employed Bayesian models with citizen science in ecology. The 84 exemplar papers are summarised with respect to the aim of the study, the statistical model/method and software employed, the data captured and the platform enlisted. Of particular interest is the acknowledgement and treatment of potential bias in the CS data. We then study how a suitable statistical model based on item response theory (IRT) can improve the quality of crowdsourced information and enable it to be used more confidently to help answer relevant ecological problems and estimate complex measures such as ecosystem health.
This model can be used to assess changes in participants' abilities over time and whether they learn with more classification opportunities.
It can also be used to identify those categories that are harder to classify and factors affecting the difficulties of the images.

Following discussion of the model, we present a case study of the classification of underwater images from the Great Barrier Reef (GBR), Australia. In the experiment, we showed coral reef images to non-expert participants and asked them to classify five broad categories of benthic communities using the Amazon Mechanical Turk platform (MTurk). The statistical approach developed in this paper allows us to estimate the most difficult benthic categories, investigate different types of difficulty, including those related to images and cameras, and detect different groups of participants, based on their abilities to perform the required tasks.
We also examine the relationship between the time that a participant needed to perform the classification and the participants' latent abilities.
%The classification dataset is made available for further research. 

The rest of the paper is structured as follows. A review of Bayesian methods for citizen science is provided in Section 2 followed by the introduction of a Bayesian item response model which addresses biases in image classification in Section 3. Section 4 details the paper's case study in the Great Barrier Reef, Australia, including a results section summarising the applied models. Finally, Section 5 contains a detailed discussion centred around the quality of citizen science data for image classification specifically and for use in ecological research.
It should be noted that we focus on images in this paper and often refer to images as `items'. Additionally, we use the term `label' to refer to the true underlying category of the target observation in any reference images. Despite dealing with the classification of images in this study, the proposed methodology can be extrapolated to other sources of citizen science-produced data such as audio recordings.

\section{Bayesian methods for citizen science}

Scientists are increasingly employing Bayesian statistical methods to overcome some of the limitations of CS data.
We conducted a scan of some of the most relevant developments, following the PRISMA guidelines (Preferred Reporting Items for Systematic Reviews and Meta-Analyses) \citep{liberati2009prisma}.
We searched the Web of Science, Scopus, and Google Scholar databases using the keywords ``citizen science'' and ``Bayes'', and filtered articles belonging to Ecology and Conservation. 
The search was restricted to articles in the English language published in peer-reviewed articles from 2010 to 30-Apr-2022. 

\begin{figure}[ht]
  \centering
  \includegraphics[width=4.5in]{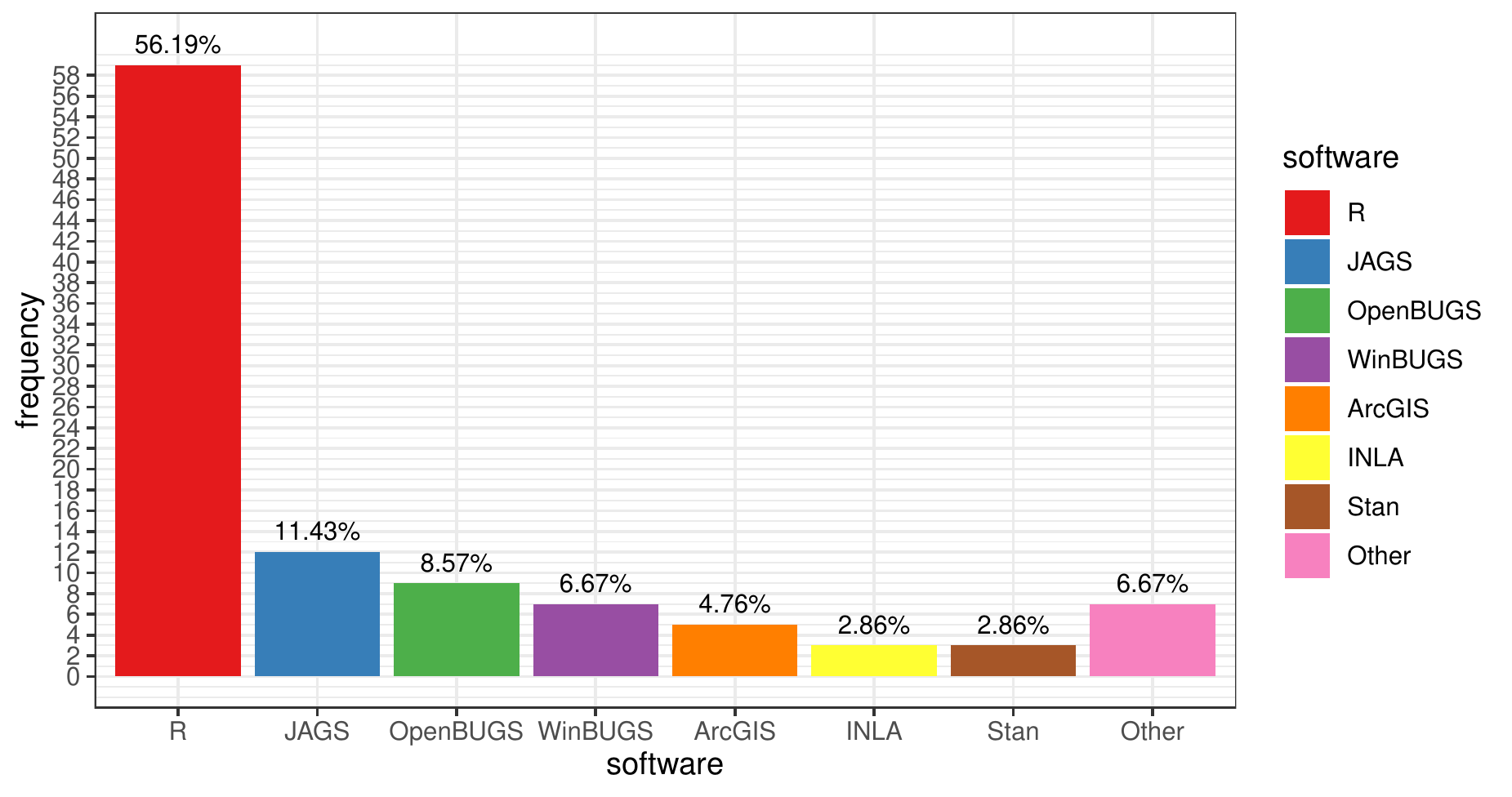}
\caption{
Reported software utilized across the reviewed papers.}
\label{fig:soft}
\end{figure}

We found a total of 84 articles meeting the inclusion criteria.
For each article, we summarised the statistical method used, the most relevant findings, the data sources, the platform/project and the software/packages used for modelling (if it was mentioned explicitly). The articles are summarized in Table~\ref{table:summ}.
In total, 36.4\% of the articles used Bayesian regression and/or hierarchical models. A comparable percentage employed Bayesian species distribution models / Bayesian occupancy models. Approximately 20\% of the models incorporate spatial and spatio-temporal dependence. Other statistical models not included in these categories accounted for 18\%.
In Fig~\ref{fig:soft} we have summarized the software used for Bayesian computation. 
Over half of the articles that specified the use of software mentioned the use of R \citep{rprog}, frequently in conjunction with dedicated software such as WinBUGS \citep{lunn2000winbugs}, OpenBUGS \citep{spiegelhalter2007openbugs}, Stan \citep{carpenter2017stan} and INLA \citep{rue2009approximate}.
Among the scanned paper, the most popular \textsf{R} packages for Bayesian inference are
\textsf{R2jags} \citep{R2jags}, \textsf{rjags} \citep{rjags}, \textsf{R2OpenBUGS}\citep{R2WinBUGS}, \textsf{rstan} \citep{RStan}, \textsf{unmarked} \citep{unmarked}, \textsf{adegenet} \citep{adegenet}, \textsf{ape}\citep{ape}, \textsf{jagsUI} \citep{jagsUI}, \textsf{R2WinBUGS} \citep{R2WinBUGS}, \textsf{rstanarm} \citep{rstanarm}.

A substantial proportion of the scanned articles deals with topics such as the estimation of the presence/absence of species or their abundance. 
Hence, many authors resort to Bayesian occupancy and species distribution models \citep[e.g.][]{WOS:000786312200001, sheard2021long, WOS:000719372700005, WOS:000686044200001, WOS:000668586900001, rodhouse2021audible}.
Ecological data is generally geo-referenced and many of the models account for spatial variation by using for instance conditional autoregressive models (CAR) \citep{arab2016spatio, croft2019modeling, dwyer2016using, purse2015landscape, santos2020correcting}, covariance matrices \citep{reich2018integrating}, Gaussian processes \citep{sicacha2020accounting}, SPDE \citep{girardello2019gaps, humphreys2019seasonal}  or simply incorporating spatially varying covariates.

One of our main aims was to explore the treatment authors give to CS-produced data. 
We found that researchers often assume that no errors or biases are present in the data, which could affect statistical inference and decision making.
For example, many CS projects in ecology use opportunistic data that are obtained without a sampling design or a professional survey, since traditional professional programs are costly and time-consuming and can fail to obtain accurate or precise estimates of infrequent species or other outcomes of interest.
Geographical, spatial recording, and preferential sampling bias may arise from opportunistically collected data, with more observations from frequently visited locations such as around roads, with irregular frequencies across time, or as a result of preferential sampling spots with certain habitat types or where specific species are more likely to be found. See the discussion in \citet{WOS:000691436000001, fournier2017combining, zhu2020migratory}.
In addition, observer bias may arise as the result of inexperience, or incorrect understanding or perception of a variable of interest. Detection bias may arise if a species or outcome of interest is missed or misidentified.

Overall, we identified several strategies that have been proposed to cope with potential bias in CS data. 
A discussion around the use of opportunistic data can be found in \citet{coron2018bayesian}.
Similarly, the temporal bias (observation efforts) arising from the seasonality of visitors to the area of study is addressed by \citet{dwyer2016using}, specifically for sightings data. 
In a spatial context, \citet{van2013occupancy} suggested using post-stratification of sites and showed how opportunistic data can produce suitable estimates when adjusting for bias produced by geographical imbalance and unequal observation efforts.
In another study, \citet{humphreys2019seasonal} included human population density to account for the fact that sightings might be more likely in areas with larger populations. 
These authors also tackled observer bias by using observation effort expressed as observer hours as a predictor in their models.

Imperfect detection is relevant, especially for data contributed by citizen scientists.
Occupancy models generally involve a component for the presence/absence and another for the detection/non-detection.
These models aim at minimizing the misclassification rate (number of false negatives and false positives). 
Variations of this models can be found in \citet{strebel2014studying}, \citet{isaac2014statistics}, \citet{petracca2018robust}  and \citet{van2013opportunistic}.
See \citet{isaac2014statistics}  for a comparison of several Bayesian occupancy models in terms of the efficiency to detect trends based on different opportunistic data collection scenarios.
Detection bias is also considered in \citet{berberich2016detection} by accounting for false-positive detection (specificity) bias when spotting red wood ant nests.
\citet{viljugrein2019method} also considered diagnostic sensitivity for the probability that an infected animal will be detected by testing.
Similarly, \citet{cumming2019point} corrected for observation effort to account for imperfect detection. 
More recently, \citet{eisma2020bayesian} suggested a measurement error model to adjust for biases and errors in reports of rainfall measurement data produced by volunteers in Nepal.

There exists a fundamental gap in the literature regarding approaches to overcome other sources of bias. This includes potential biases resulting from the misclassification errors arising during the classification of objects on images by citizen scientists. 
We discuss this issue in the next section.

\section{Addressing citizen science bias in image classification} 

In this section, we first discuss majority voting algorithms that are frequently used to estimate the true labels in manual image classification.
This is followed by the introduction of a Bayesian item response model that is used to weight the evidence produced by citizen scientists.

\subsection{Majority vote (MV)}

We consider a set of images $j = 1, 2, \cdots, J$, each composed of $k = 1, 2, \cdots, K$ elicitation points selected using a spatially balanced random sampling approach. 
We deal with a binary classification task, in which a participant $i$ is asked whether a category (e.g., coral) is present on a point belonging to a given image.
Let $Y_{ijk}$ be the answer of the participant $i$ for the $k^{\textrm{th}}$ point from the $j^{\textrm{th}}$ image.

\begin{equation} 
\label{eq:pijk}
Y_{ijk} = \left \{ \begin{matrix}
1 \:\:\textrm{if the participant considered the category to be present}  \\ 
0  \:\:\textrm{Otherwise}.
\end{matrix} \right.
\end{equation}

By design, multiple participants classify the same point $k$. 
Based on these answers, we obtain the majority vote by aggregating the answers, so that the category with the highest proportion of votes wins the vote, which is the mode.

\begin{equation} 
%\begin{align}
\label{eq:pijl}
\hat{Y}_{jk} = \left \{ 
\begin{matrix}
1 \:\:\textrm{if proportion of votes with ``category present''  $>$ 0.5} \\ %i.e. \sum_{i=1}^{I}z_{jk}/(I) 
0 \:\:\textrm{Otherwise.}
\end{matrix} \right .
%\end{align}
\end{equation}

In general, this approach performs poorly for difficult tasks.
This is because only expert participants are likely to respond correctly, and they can be outvoted by beginners  \citep{raykar2010learning}.
A variation of this method is obtained using a weighted majority voting (WMV) which has been discussed, among others, by \citet{littlestone1989weighted, lintott2008galaxy, hines2015aggregating}.
In this method, each participant has a vote proportional to some weights, based on their knowledge, skills or past performance.

\subsection{Bayesian IRT model}

The estimation of participants' abilities in crowdsourced data has been widely discussed in the literature \citep{whitehill2009whose, welinder2010online, paun2018comparing}. We develop a Bayesian item response model with the aim of informing a weighted consensus voting approach.
Reiterating for completeness the notation introduced earlier, let the binary response variable $Y_{ijk}$ represent whether a question associated with the $k^\textrm{th}$ point ($k = 1,\cdots,K$) on the $j^\textrm{th}$ image ($j = 1,\cdots,J$) taken using the $l^{\textrm{th}}$ camera
is correctly answered or not by the $i ^\textrm{th}$ participant ($i = 1,\cdots,I$). 
We assume that $Y_{ijkl}$ follows a Bernoulli distribution with parameter $p_{ijkl}$

\begin{equation}
Y_{ijkl} \sim \textrm{Bern} \left(p_{ijkl} \right).
\label {eq:eq1}
\end{equation}

We use an extension of the item response model, namely the linear logistic test model (LLTM), which is formulated as follows

\begin{equation} \label{eq:long}
	p_{ijkl} = \eta_k + \left(1 - \eta_k \right)\frac{1}{1+\textrm{exp}\left \{ -\alpha_k\left ( \theta_i - \beta_k - \beta_l\right ) \right \}},
\end{equation}

\noindent where $\beta_k$ and $\beta_l$ are difficulties associated with the point and the camera. 
The parameter $\theta_i$ represents the ability of the participant. The ability of an average participant can be anchored by setting it equal to zero to avoid identifiability issues with the model.
Additionally, $\alpha_k$ gives the slope of the logistic curve and $\eta_k$ is a pseudo-guessing associated with the point, indicating the probability of answering correctly due to guessing. 
We use Bayesian inference and therefore we need to define prior distributions for the parameters of interest in Eq\ref{eq:long}.

{ \scriptsize
\begin{flalign*} \label{eq:priors}
		\theta_{i}&\sim \mathcal{N}\left(0,\sigma_{\theta}\right)   && \text{\# hierarchical prior on the abilities}\\
		\sigma_{\theta}&\sim \textrm{uniform}\left(0,10\right)  && \text{\# flat prior on a weakly informative range for the s.d. of the users' abilities}\\
		\beta_{k}&\sim \mathcal{N}\left(\mu_{b_k},\sigma_{b_k}\right)  && \text{\# hierarchical prior on the item difficulties}\\	
    \mu_{\beta_{k}}&\sim \mathcal{N}\left(0,5\right) && \text{\# weakly informative prior for the mean of the item difficulties }\\
		\sigma_{\beta_{k}}&\sim \textrm{Cauchy}\left(0,5\right)T(0,\infty)  && \text{\# informative prior for the sd of the item difficulty, allowing for substantially complex tasks }\\
		\beta_{l}&\sim \mathcal{N}\left(\mu_{b},\sigma_{b_{l}}\right)  && \text{\# hierarchical prior on the camera difficulties}\\
		\mu_{\beta_{l}}&\sim \mathcal{N}\left(0,5\right) && \text{\# weakly informative prior for the mean of camera difficulty }\\
		\sigma_{\beta_{l}}&\sim \textrm{Cauchy}\left(0,5\right)T(0,\infty)  && \text{\# informative prior for the sd of the camera difficulties }\\
		\alpha_k &\sim \mathcal{N}\left(1,\sigma_{\alpha}\right) && \text{\# normal prior with mean 1 on the slope}\\
		\sigma_{\alpha}&\sim \textrm{Cauchy}\left(0,5\right)T(0,\infty)  && \text{\# half Cauchy prior on the slope sd truncated at 0}\\
		\eta_k & \sim \textrm{beta}\left(1,5\right)   &&  \text{\# weakly informative prior on the pseudoguessing}
%\end{align*} 
	%\end{equation}
\end{flalign*}
}

\subsubsection*{Changes in ability}

Several authors have suggested the implementation of dynamic item response models that account for temporal variation in the answers,
under the principle that subjects' abilities change with time as a learning curve \citep[e.g.][]{wang2013bayesian}.
To capture the learning in the process, we incorporated a temporally dependent component to the model
\begin{equation} 
	p_{ijklt} = \eta_k + \left(1 - \eta_k \right)\frac{1}{1+\textrm{exp}\left \{ -\alpha_k\left ( \phi_t + \theta_i - \beta_k - \beta_l\right ) \right \}},
	%p_{ijlt} = \eta_j + \left(1 - \eta_j \right)\frac{1}{1+\textrm{exp}\left \{ -\alpha_j \left ( \theta_i + \phi_t - \beta_jI_{j} - \beta_lI_{l}\right ) \right \}}
\label{eq:eqtime}
\end{equation}

\noindent where $\phi_t$ is a common learning measure that captures the change in abilities according to the daily occasions that participants performed classifications $t = 1, 2, \cdots 15$.
For participant $i$, $t=1$ represents the first classification day,  $t=2$ the second, and so on.

\subsection*{Consensus based on a Bayesian item response model}

The model described above is set in the context of a broader workflow, described as follows:

\begin{enumerate}
	\item Produce a representative set of gold standard images e.g. 33\% of the total number of images. In this set of images the true labels or answers will be obtained from expert elicitation or other another suitable method. 
	Images in this set will be scored by most of the participants. %This could be the first images presented to the participants.
	\item Fit an item response model and obtain estimates of the participants' abilities accounting for difficulties, guessing, etc.
	\item For a weighted consensus, derive a weight for each participant proportional to their estimated ability using the posterior mean. We can compute the weights using $w_i = exp(\theta_i) / \sum_{i=1}^{I}(exp(\theta_i))$. 
 Alternatively, a fully Bayesian framework can be employed by using the draws from the posterior distribution of $\theta_i$ to compute the distribution of $w_i$.
    \item Perform a weighted consensus vote to estimate the labels.  Since our response variable is binary, the category with the largest $\sum_{k=1}^{K} w_i$ wins the weighted vote.
\end{enumerate}

\section{Case study: Classification of images from coral reefs}

{ \footnotesize\emph{``How inappropriate to call this planet Earth when it is quite clearly Ocean.''} - Arthur C. Clarke}\\

The Great Barrier Reef (GBR) is located on Australia's north eastern coast and is among the largest and most complex ecosystems in the world \citep{authority2009great}.
Two impacts of climate change, including the increasing frequency of reef bleaching events and intensity of cyclones, are negatively affecting this ecosystem causing an unprecedented decline in the prevalence of hard corals \citep{hughes2017coral, de201227, ainsworth2016climate, vercelloni2020forecasting}. 
Estimation and assessment of this decline are difficult and expensive to quantify using traditional marine surveys considering the size of the GBR and the speed of decline \citep{gonzalez2020monitoring}. For this reason, some researchers are harnessing the strength of citizen science to produce estimates of reef-health indicators across large spatial and temporal scales \citep{peterson2020monitoring, santos2020correcting}. 
For instance, \citet{santos2020correcting}, utilized a spatial misclassification model, taking into account the proficiency of the participant in terms of sensitivity and specificity to account for bias in the data. 
Information from these models can then be used by reef managers and scientists to make data-enabled management decisions and inform future research.

We performed an experiment using Amazon Mechanical Turk (\url{https://www.mturk.com/}) to assess the feasibility of using crowdsourced data for the estimation of hard coral cover, represented as the two-dimensional proportion of the seafloor covered in hard corals. Hard corals play an important role in reef ecosystems; their hard skeletons provide habitat for many organisms and they are vulnerable to a range of impacts that accumulate with climate change \citep{hughes2017coral}. 
The dataset used in the study consisted of 514 geotagged images obtained from the XL Catlin Seaview Survey \citep{gonzalez2014catlin} and the University of Queensland's Remote Sensing Research Centre \citep{roelfsema2018coral}, which we used to assess the participants' abilities to identify hard corals. In practice, coral cover estimates from images are often based on a subset of classification points, rather than the whole image \citep{thompson2016marine, sweatman2005long}. In these two programs, 40 to 50 spatially balanced, random classification points were selected on each image and classified by coral reef scientists \citep{gonzalez2020monitoring,roelfsema2021benthic}. We consider the classifications from the reef scientists as a gold standard (i.e. the ground truth).

We engaged participants and provided instructions in an 11-page training document \url{https://github.com/EdgarSantos-Fernandez/reef_misclassification/HelpGuide_MTurk20200203.pdf}, describing how to identify the different benthic categories, which included: \emph{hard} and \emph{soft} corals, \emph{algae}, \emph{sand}, \emph{water} and \emph{other}. Participants were also given the option to select \emph{unsure} if they were uncertain about which category to select.
Several image classification examples were included in this guide and the differences between commonly misclassified benthic groups were highlighted \url{https://www.virtualreef.org.au/wp-content/uploads/VirtualReefDiver-Classification-HelpGuide-Part2.pdf}. 

After studying the training document, a qualification task was used to assess the proficiency of the participants to accurately complete the task. More specifically, the participants were shown five images containing one classification point each and asked to select the correct class from the five possible choices. The qualification was granted to those scoring at least three correct classifications out of five.

We designed a sampling protocol to select images representative of the GBR in terms of community composition (proportion of hard and soft corals, algae, and sand) and camera types (Canon, Lumix,
Olympus, Sony, and Nikon). This produced a dataset composed of 514 images.
We produced 514 human intelligence tasks (HITS) with a maximum number of 70 assignments per HIT (i.e. maximum number of times each image can be classified). 

\begin{figure}
  \centering
  \includegraphics[width=1.05\linewidth]{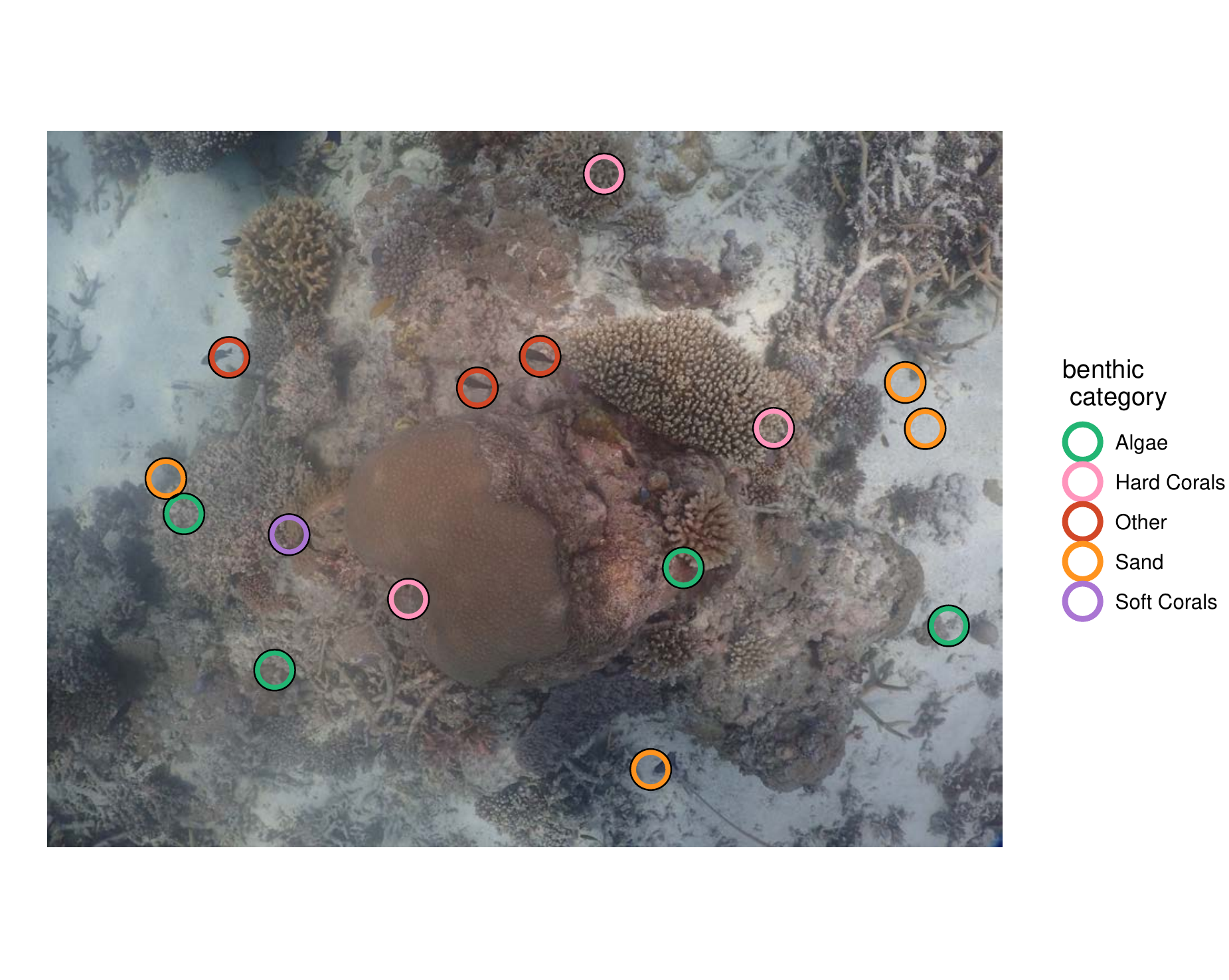}
  \label{fig:sfig2}
\caption{Example of an underwater image from the Great Barrier Reef, Australia used for classification.
 Participants were asked to classify what they observed within the points into multiple categories.  }
\label{fig:reefimages}
\end{figure}

Images were randomly assigned to the participants.
We were concerned that classifying 40 or 50 points per image was too time consuming and that it would reduce participation. In addition, previous research has shown that accurate estimates of coral cover can be obtained with approximately 10 points \citet{beijbom2015towards}. Therefore, we asked participants to classify 15 classification points on each image, randomly selected from the 40-50 points previously classified by reef scientists. See the example in Fig.\ref{fig:reefimages}.  
Participants were required to select a classification category for all of the points before submitting the classification.
Every assignment (i.e. an image) was expected to take approximately one minute to complete.
The payment was set to 0.10 USD per image and participants reported earning more than the federal minimum wage in the United States (\$7.25 per hour) for their contributions.
We monitored the quality of the classifications to prevent low-quality participants from contributing.

\subsection{Performance measures}

Our category of interest in the analyses is {\it hard corals}. We used a suite of performance measures to describe the ability of the participants, which are based on the true positive (TP), true negative (TN), false positive (FP), and false negative (FN). 
In this context, TP are the points classified as hard coral given that the point is truly occupied by hard corals. A TN occurs when the point is correctly classified as something other than hard coral when there is no hard coral present. Similarly, FP represents points classified as hard coral when hard coral is absent, while FN occurs when hard coral is incorrectly classified as something else. This information was then used to generate other performance measures such as:
\begin{itemize}
\item Sensitivity: measures the ability of a participant to identify a category when it is present: $se = TP/(TP + FN)$.
\item Specificity: measures the ability of a participant to correctly identify a category when it is absent:  $sp = TN/(TN + FP)$.
\item Classification accuracy: the proportion of correctly identified classification points: $acc = (TP + TN)/(TP + TN + FN+ FP)$.
\item Precision: the probability of correctly classifying points that truly contain the category hard coral divided by the total number of points classified as containing it $pre = TP/(TP + FP)$.
\item Matthews correlation coefficient (MCC) \citep{matthews1975comparison}: a measure of the effectiveness of the classifier using all the elements of the confusion matrix. 
\item positive likelihood ratio ($lr_+$): gives the number of true positives for every false positive.
\item negative likelihood ratio ($lr_-$): measures the number of false negatives for every true negative.
\end{itemize}

We fitted the Bayesian item response model given in Eq \ref{eq:eqtime}. 
We extracted the posterior distribution of the abilities ($\theta_i$) and clustered subjects. 
We then used these clusters to construct several model variations based on majority voting, restricted to experts or experts/experienced subjects.
Three replicates were generated for different gold standard proportions and the results were averaged out.

\subsection{Results}

The data contributed by participants were aggregated using the five different methods.
First, we considered the \emph{raw} estimates obtained directly from the participants' classifications (i.e. raw), without any grouping or consensus applied (Table \ref{table:model_results}). 
The columns in this table represent the number of classification points ($n$) and the performance measures.

We considered a traditional consensus and weighted variation based on item response model from Eq.\ref{eq:eqtime}. The robustness of the results were assessed for the proposed consensus method using different proportions of images where the truth was known (10\%, 20\%, 33\% and 50\%); noting that images were randomly selected without replacement.

Using the raw data without combining the subjects' answers produced relatively low-quality performance compared to other aggregation methods, $se = 0.642$ and $sp = 0.618$ with an accuracy of 0.626 and $MCC = 0.246$. On average, the subjects identified 1.682 TP hard coral points for every FP hard coral classification.
Using a traditional consensus approach combining all of the participants' responses substantially increased the performance measures compared to the raw data method e.g. $se = 0.825$ and $MCC = 0.487$.

\begin{figure}[hb]
	\centerline{\includegraphics[width=3.4in]{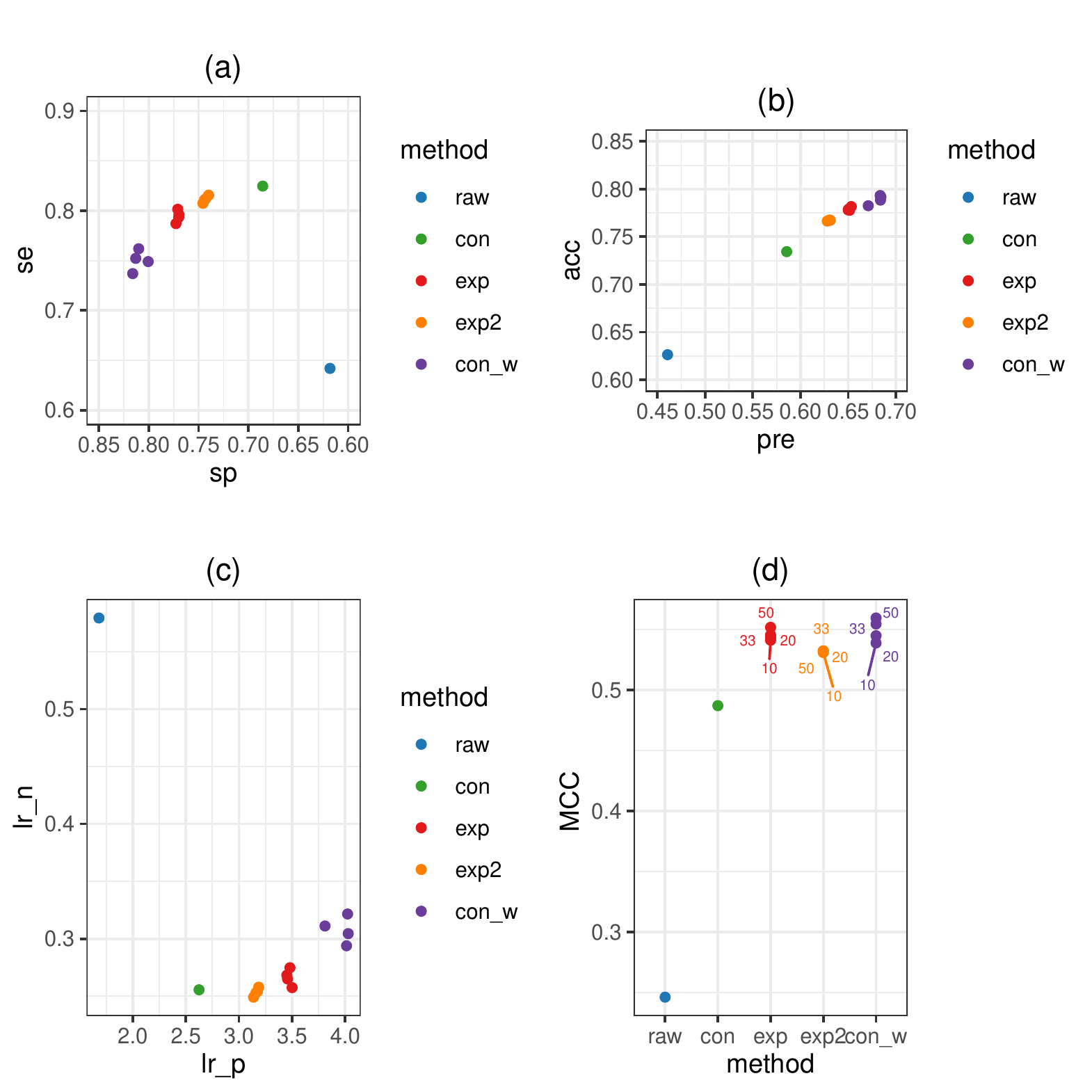}}
	\caption{Comparison of the raw performance measures and those obtained from different methods: consensus (con), 
	item response consensus based on experts/experienced (exp) \& experts participants (exp2), and
	using weighted consensus (con\_w). 
	(a) sensitivity ($se$) vs specificity ($sp$),
	(b) accuracy ($acc$) vs precision ($pre$),
	(c) negative likelihood ratio ($lr_n$) vs positive likelihood ratio ($lr_p$),
	(d) Matthews correlation coefficient ($MCC$).
	} 
	\label{fig:methods_comparison}
\end{figure}

The majority voting methods using a selection of the participants outperform ($MCC > 0.53$) in all the variants explored.
The ratio of TP/FP points identified under these methods is above 3.
However, the likelihood ratio (FN/TN) was similar to the one obtained using the consensus approach.
Fig \ref{fig:methods_comparison} compares the statistical performance measures. In the case of the MV methods, we show four values per method representing the proportion of images in the gold standard set (10, 20, 33 and 50\%). 
The last method (weighted variation) performed well compared to the approach based on experts and experienced, achieving marginal improvements in $acc$ and $pre$.     
We found that the item response model captures well the abilities of the subjects, even with a small training dataset.
This indicates that a minimal training set (e.g. 50 images) is enough to cluster subjects based on proficiency (i.e. expert, experienced, etc.).

\begin{figure*}
%produced with file 51 mturk.R
\centerline{\includegraphics[width=5in]{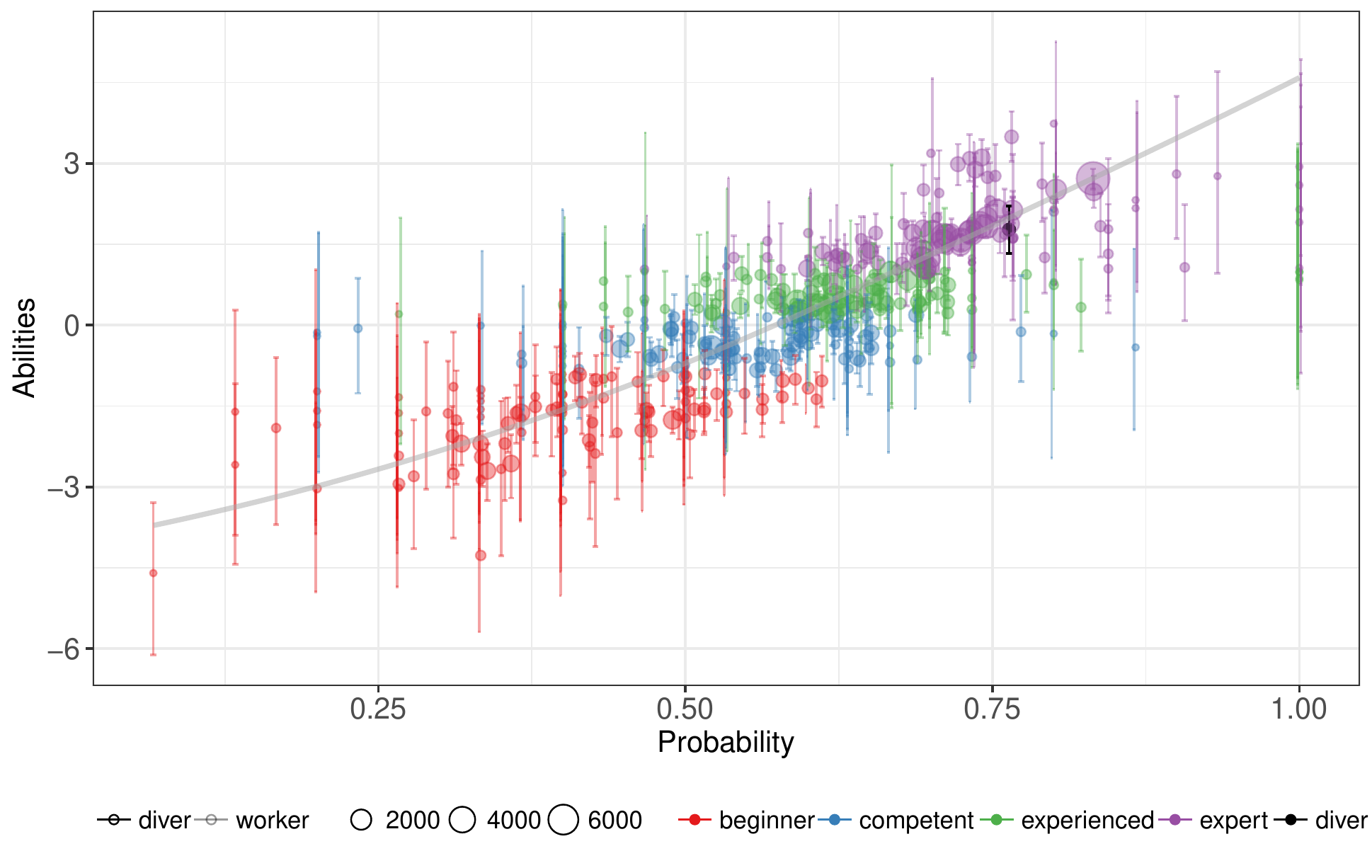}}
	\caption{Abilities posterior estimates for four groups of participants (beginner, competent, experienced and experts) using a gold standard dataset ($n = 171$) and 95\% highest density interval as a function of the proportion of correct answers. 
	The size of the dot represents the number of points classified. The black dot represents a reference diver that engaged in the projects as a participant.} 
	\label{fig:Fig_abil}
\end{figure*}

Four groups of participants were obtained from the quantiles of the posterior means of the abilities (Figure \ref{fig:Fig_abil}): {\it beginners}, {\it competent}, {\it experienced} and {\it experts}. 
The vertical axis gives the latent ability score ($\theta_i$) and the x-axis the proportion of correctly classified points.
Skilful participants have large ability score values. The size of the point gives the number of classification points and indicates the engagement in the project.
The vertical bar is the 95\% posterior highest density interval and represents the dispersion around the posterior ability estimate.
In black, we represent a reference participant who self-identified with diving experience. 
This participant falls within the expert category, yet remarkably it was not the best-performing one.

\begin{figure}
%PRODUCED BY file 53 abil
		\centerline{\includegraphics[width=3.0in]{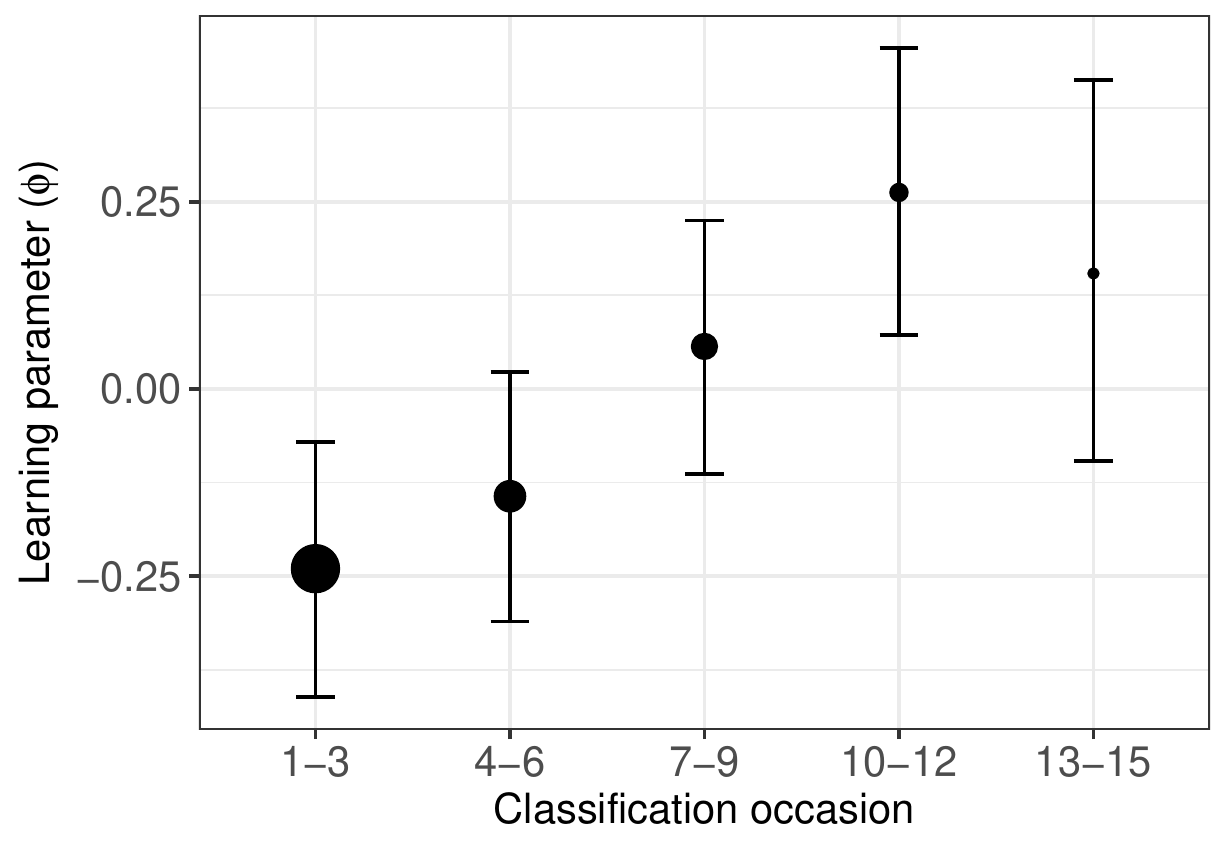}}
	\caption{Posterior estimates of the learning parameter $\phi$ and 95\% highest density interval as a function of the daily classification occasion.
	These estimates were obtained fitting the whole dataset ($n=514$ images).
	The size of the dot is relative to the number of points classified.
	The error bars represent the 95\% highest density interval of $\phi$.
	%The red line represent the fit of a linear model to the posterior mean of $\phi$. The slope is significantly different from 0 (p = 0.019).
	} 
	\label{fig:Fig_leaning}
\end{figure}

In Fig~\ref{fig:Fig_leaning} we assess the dynamics of the rate of learning as the participants increased daily classification occasions (i.e. they work on the task a second, third, fourth time, etc). 
Fitting a linear regression to the posterior means at the occasion $t$ produced a slope significantly different from 0 (p-value = 0.019), which indicates that the participants' skills increase with participation and they become better at classifying the points.
After controlling for pseudo-guessing and discrimination, the average participant increased their probability of correct classification by approximately 4\% after five occasions, and 8\% and 12\% in the $10^{th}$ and $15^{th}$ occasions, respectively. 

\begin{figure}
	\centerline{\includegraphics[width=4.0in]{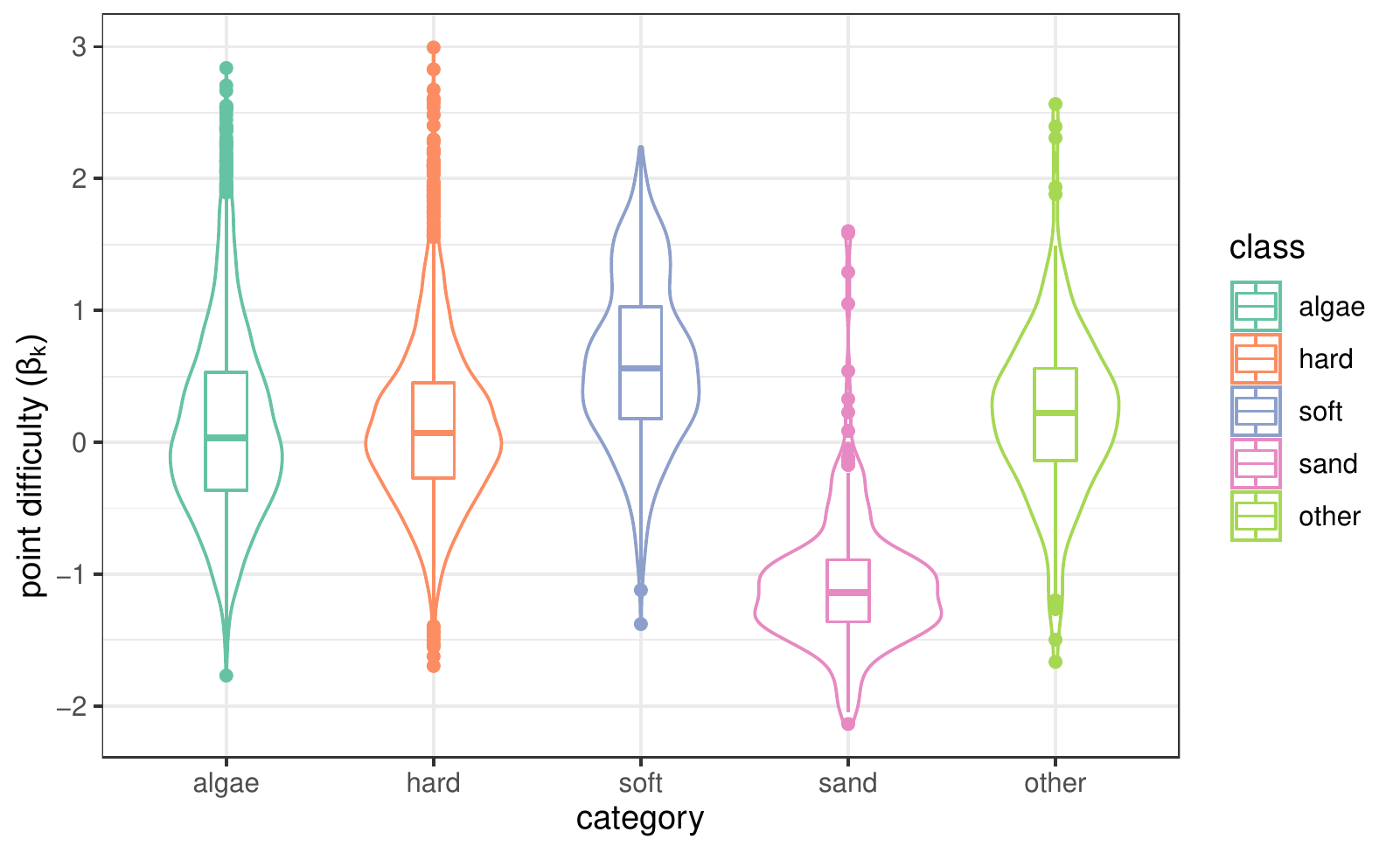}}
	\caption{Violin/box plots with difficulty parameters of the points on images ($\beta_k$) associated with the benthic categories}. \label{fig:species_difficulties}
\end{figure}

\begin{figure}
	\centerline{\includegraphics[width=3.5in]{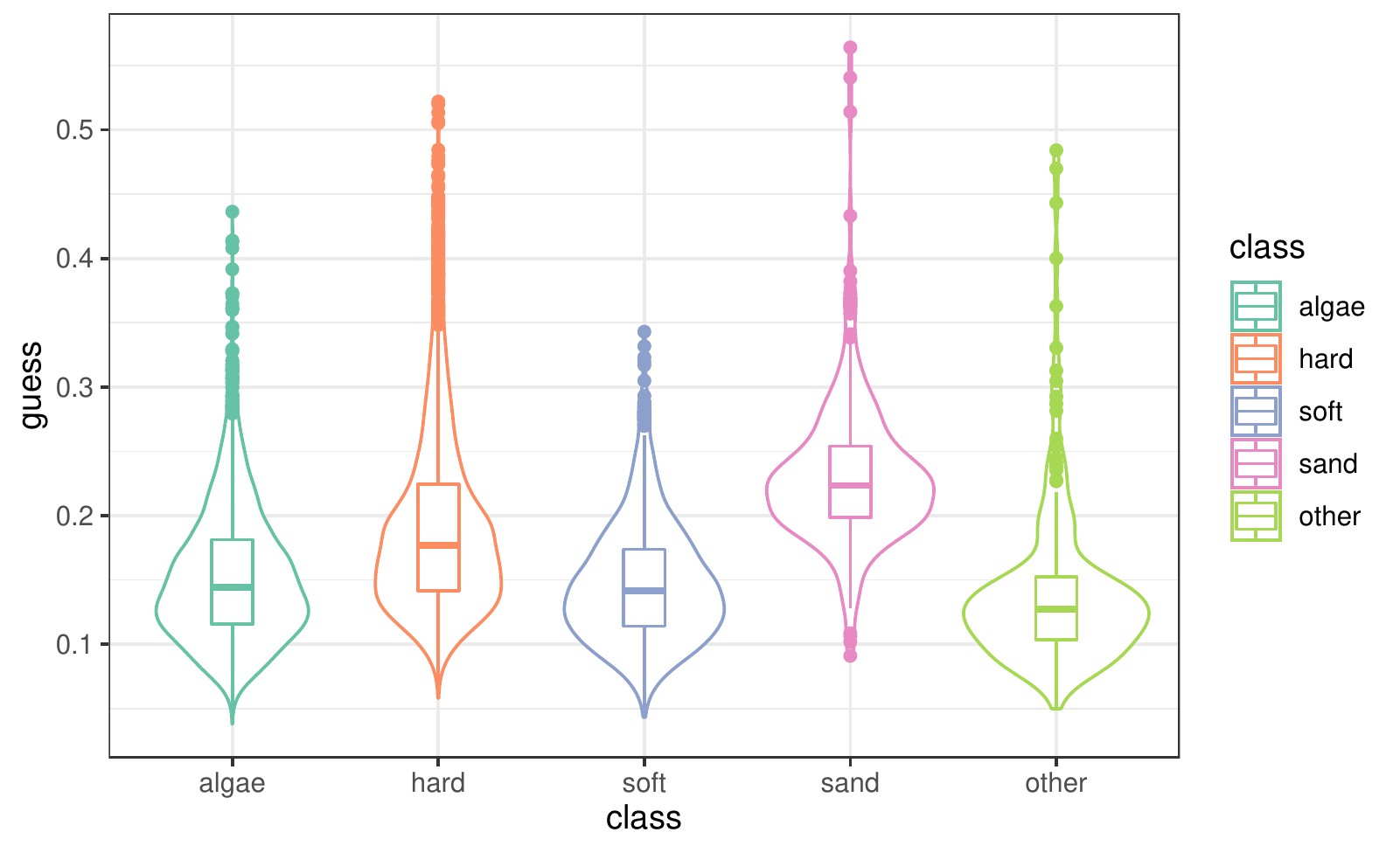}}
	\caption{Violin/box plots of the posterior estimates of the pseudoguessing parameter associated with the benthic categories.}
	\label{fig:species_guess}
\end{figure}

There were also differences in the difficulty of classifying each of the classes (Fig~\ref{fig:species_difficulties}). The results showed that the soft coral category was the hardest to identify in the images since they are frequently misclassified as hard corals.
As expected, points containing sand had the highest chances of correct classification.   
This category also exhibited the highest likelihood of being correctly identified by chance, as evidenced by the posterior distribution of the pseudo-guessing parameter (Fig~\ref{fig:species_guess}). 
%This category had also the highest probability of identification by chance given by the posterior distribution of the pseudo-guessing parameter (Fig~\ref{fig:species_guess}).

\begin{figure}
% produced by 51 mturk.R
	\centerline{\includegraphics[width=3.5in]{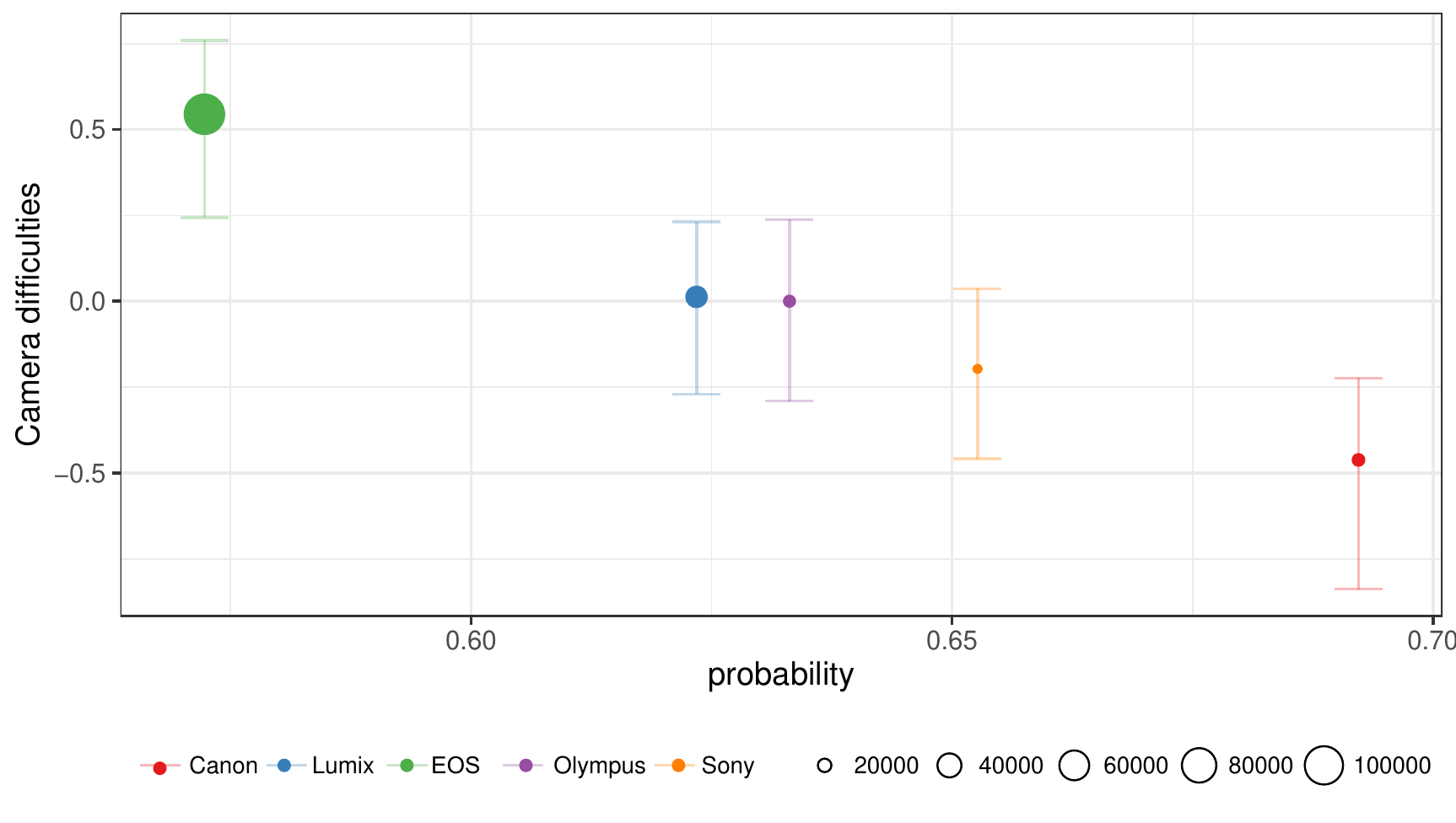}}
	\caption{Posterior estimates of the difficulties and 95\% highest density interval of the five cameras used on the study as a function of the proportion of correct answers. 
	The size of the dot represents the number of points classified in images taken from each camera.}  
	\label{fig:camera}
\end{figure}

Substantial differences were found among the cameras used to take the images. Images taken by the Nikon camera represented the vast majority and they were substantially more difficult (Fig~\ref{fig:camera}). This camera was used to take images in the northern section of the GBR on a unique habitat as part of the XL Catlin Seaview survey \citep{gonzalez2014catlin}. Images from the Canon camera, taken as part of the Heron survey across different habitats, were easier to classify \citep{roelfsema2010calibration}. 
However, the difficulty associated with the camera might be confounded by the regions where images were taken (North versus South of the GBR and habitats), where biodiversity differs.

\subsubsection{Classification time and quality}

We evaluated the feasibility of using classification time as a straightforward indicator of classification quality.
This is relevant in the context where participants are getting paid for completed images and thus trying to maximize their effort and it could be used as a straightforward way of detecting low-performing respondents. 

\begin{figure}
% file 51 mturk
\centerline{\includegraphics[width=3.5in]{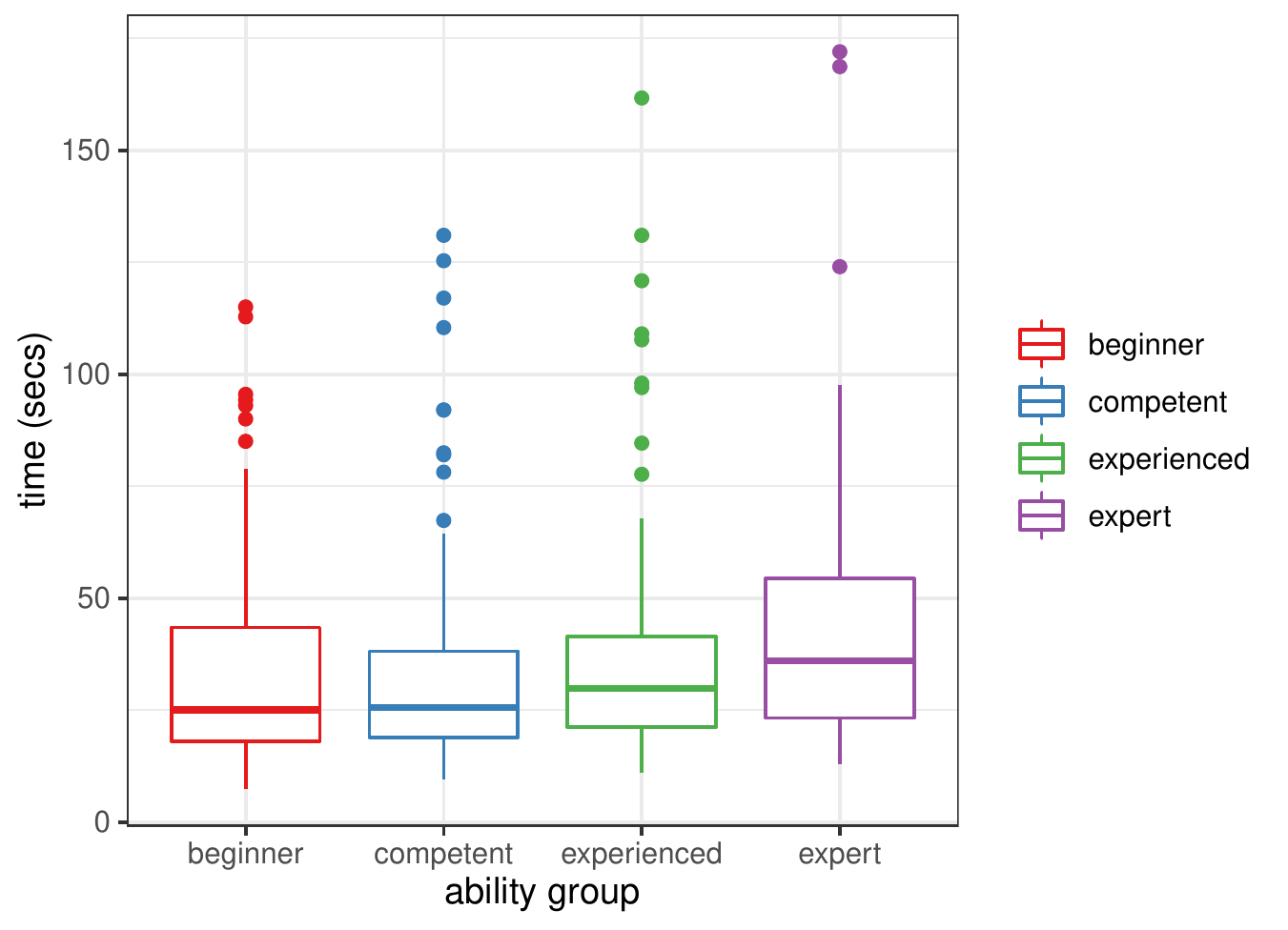}}
	\caption{Box plots of the classification time as a function of the participants ability groups.}  
	\label{fig:timeVSability}
\end{figure}

The boxplots in Fig~\ref{fig:timeVSability} show the distributions of the classification times for each of the participant groups.
We note that those participants identified as expert  seem to require significantly higher median classification times compared to those in the beginner and competent groups as shown in Table~\ref{table:wilcox}.  
The comparison was made using the non-parametric Wilcoxon test, with an alternative hypothesis that the sampled elements from the group in the rows had substantially \emph{greater} mean rank values than those in the columns (Table~\ref{table:wilcox}).
However, due to the substantial overlap between these distributions and variability in the data, these results should be interpreted with caution.

\section{Discussion} 

\subsection{Improving trust in citizen science data}

Citizen science (CS) has become an essential information source in many domains. However, the validity of research outputs using these data sources is often questioned, especially when participants with varying skills are involved in challenging tasks. 
The lack of trust in this new type of data hampers the full potential of CS programs to support management and data-driven decision-making. 
Our scan of recently published papers highlighted the broad use of CS data in addressing ecological questions such as mapping species abundance and understanding their associated drivers through time. 
Some, but not all, of the studies recognized the potential inherent bias in CS data, acknowledging its ecological implications and adapting methodological approaches accordingly.
Notwithstanding these efforts, further work is required to advance current statistical practice with CS data.

\subsection{Improving trust in citizen science data for image classification}

Our case study focused on the classification of objects in images and described a method that weights the evidence to produce the true latent labels, which are used to predict the health of coral populations along the Great Barrier Reef, Australia. 
This was achieved via estimation of participants' abilities after accounting for factors such as image difficulty.

Applications based on image-based classification data are drawing increasing attention in many domains. 
It is therefore natural that there will be further opportunities to learn from the contribution of participants.
However, new statistical methods are required if these opportunities are to be meaningfully realised.
For example, the identification of benthic categories on images is currently deemed to be too challenging for most citizen scientists, who have little knowledge of what a hard coral looks like. Similarly, not all participants have the same commitment and skills and they engage differently in CS programs. This is critical when dealing with CS data and statistical models need to weigh the evidence, based on these factors.

Obtaining the true classes using marine biologists or expert elicitation is expensive when large numbers of images need to be classified. Our results show that an item response model is a viable option when there are budget constraints.
The item response modelling framework provides an effective way to assess participants' skills and cluster/group them by the level of expertise. 
This approach allows flexibility for the CS programs to select data and perform various assessments along the way \citep{santos2021}. 
For example, our case study demonstrates increasing learning of the participants with time. 
This insight showcases the importance of retention in CS programs as people naturally learn, even with complicated tasks. 
These models also allowed us to identify careless or low-skilled respondents and detect software bots, who generally fall into the beginners' category.
Responses from these participants are generally messy; therefore our voting algorithm does not include them. In our experience, this step was critical to achieving good classification performance.    
   
We also showed that multiple factors affect the difficulty of the task, including the underlying category on the images and the camera type. Identifying those categories and images that produce greater misclassification errors is critical to producing useful training and qualification materials.
Identifying and combining expert responses is a suitable solution and this can be done even with a reduced gold standard dataset.

Citizen science project managers can benefit from the approach in many ways: 
(1) Clustering participants allows weighting the evidence; (2) the ability to identify beginners means that they can be asked to re-qualify before contributing additional data; (3) gamification, such as leaderboards, can be constructed using the latent ability values and the number of classifications; and (4) a priori knowledge about which images are the most difficult could be used to assign them to the most skillful participants.

Currently, paid platforms such as Amazon Mechanical Turk do not consider the contribution of the participants or their expertise. Instead, whoever requests the job can refuse payment for poor-quality work.
A system based on ability scores, as developed in this research, constitutes a better approach to compensate participants and is more effective than the current binary system. 
We found that for easy classification tasks, with broad evenness in the responses of the participants, most approaches will perform relatively well. 
However, when the task is difficult, aggregating the answers of the participants using simple consensus tends to have poor performance. 
Using instead a weighted majority voting approach improves performance outcomes.

\subsection{Improving trust in citizen science data for ecology}

The literature scan showed that most existing applications involve land-based ecosystem datasets.
A substantial gap exists in the use of CS for marine ecology, especially in coral reef studies.
Reef CS programs mostly exist to empower people and increase awareness, but few of them engage with robust and regular assessments of data quality which, in turn, reduce the trust of scientists and managers in using information collected by non-experts. 
However, marine research is evolving and collaborative CS programs such as Virtual Reef Diver combines modern statistical modelling and coral reef ecology to improve the integrity of CS data for decision-making. 
These methods can help marine ecology realise its full potential and contribute preservation of the health of the Great Barrier Reef by harnessing the strong engagement of several groups, including the online community. 
The valuable feedback we received in the study allowed us to improve project design, training, and compensation processes.  
An important feedback loop is also closed when measures of performance from the model are shared with the participants.

Guidelines when analysing new data sources such as CS data are essential to increase the trust among the scientific community. 
Our study focuses on Bayesian methods to analyse CS data, we acknowledge that other quantitative approaches not considered here, especially from the general crowdsourcing literature, can also add important value to CS projects. 
The item response model or similar statistical techniques should be systematically applied to first quantify and assess the quality of CS data and second support choices of a further analytical framework for ecological purposes. 
For example, insights from the model can be combined into species distribution models to weight observations according to participant skills or other parameters of interest before estimations of occupancy probability.

\section*{Data availability statement} 

The dataset used in the case study can be found in the repository:  
\url{https://github.com/EdgarSantos-Fernandez/reef}.

\section*{Acknowledgement}
This research was supported by the Australian Research Council (ARC) Laureate Fellowship Program under the project ``Bayesian Learning for Decision Making in the Big Data Era'' (ID: FL150100150) and the Centre of Excellence for Mathematical and Statistical Frontiers (ACEMS). 
We thank the editor and the reviewers for their valuable feedback and constructive suggestions.
Thanks to the members of the VRD team (\url{https://www.virtualreef.org.au/about/}).
We also like to thank all the participants who contributed to the classification of images.
Ethical approval was granted for the collection of this data by the Research Ethics Advisory Team, Queensland University of Technology (QUT). Approval Number: 1600000830.
Computations were performed through the QUT High Performance Computing (HPC) infrastructure.

\par\null

%You can cite bibliographic entries easily in Authorea. For example, here
%are a couple of citations~\cite{Cavalleri_2016,Gregory_2015}. By default citations areformatted according to a format which INSR accepts~\cite{Meskine_2019}

\par\null

\selectlanguage{english}
%\FloatBarrier
%\bibliography{bibliography/converted_to_latex.bib}
\bibliography{ref}

\clearpage
\section{Tables}

\begin{table*}[hb]
\centering
\caption{{\bf Method 1}. Performance measures obtained from the participants' classifications using raw data, and using classic consensus and item response consensus estimates. 
We considered several proportions of images where the ground truth is known (10, 20, 33 and 50\%).
%The Matthews correlation coefficient is represented in last column (MCC).
} 
%\vspace{-0.5cm}
\label{table:model_results}
\scalebox{0.76}{
\begin{tabular}{lrrrrrrrrrrrr}
  \hline
  method & n & TP & FP & TN & FN & $se$ & $sp$ & $acc$ & $pre$ & $MCC$ & $lr_+$ & $lr_-$ \\ %Matthews correlation coefficient%
  \hline

 raw &   614,160 & 132,752 & 155,482 & 251,849 & 74,077 & 0.642 & 0.618 & 0.626 & 0.461 & 0.246 & 1.682 & 0.579 \\ \hline
 consensus &  23,488 & 6,779 & 4,798 & 10,470 & 1,441 & 0.825 & 0.686 & 0.734 & 0.586 & 0.487 & 2.624 & 0.256  \\ \hline
 experts, GS:10\%,$n=51$  & 23,488 & 6,468 & 3,468 & 11,794 & 1,750 & 0.787 & 0.773 & 0.778 & 0.652 & 0.541 & 3.480 & 0.275 \\ 
 experts, GS:20\%,$n=102$  & 23,488 & 6,520 & 3,510 & 11,752 & 1,699 & 0.793 & 0.770 & 0.778 & 0.650 & 0.543 & 3.451 & 0.268 \\ 
 experts, GS:33\%,$n=171$  & 23,488 & 6,541 & 3,513 & 11,747 & 1,677 & 0.796 & 0.770 & 0.779 & 0.651 & 0.545 & 3.457 & 0.265 \\ 
 experts, GS:50\%,$n=257$  & 23,488 & 6,588 & 3,495 & 11,767 & 1,632 & 0.801 & 0.771 & 0.782 & 0.653 & 0.552 & 3.500 & 0.258 \\ \hline

 experts/experienced, GS:10\%,$n=51$ & 23,488 & 6,637 & 3,883 & 11,385 & 1,583 & 0.807 & 0.746 & 0.767 & 0.631 & 0.531 & 3.186 & 0.258  \\ 
 experts/experienced, GS:20\%,$n=102$  & 23,488 & 6,665 & 3,906 & 11,362 & 1,555 & 0.811 & 0.744 & 0.767 & 0.631 & 0.532 & 3.174 & 0.254 \\ 
 experts/experienced, GS:33\%,$n=171$  & 23,488 & 6,704 & 3,969 & 11,299 & 1,516 & 0.816 & 0.740 & 0.766 & 0.628 & 0.532 & 3.137 & 0.249 \\ 
 experts/experienced, GS:50\%,$n=257$  & 23,488 & 6,675 & 3,925 & 11,343 & 1,545 & 0.812 & 0.743 & 0.767 & 0.630 & 0.532 & 3.160 & 0.253  \\ \hline

weighted, exp/exp,GS:10\%,$n=51$ & 23,488 & 6,155 & 3,043 & 12,219 & 2,063 & 0.749 & 0.801 & 0.783 & 0.671 & 0.539 & 3.810 & 0.311  \\ 
weighted, exp/exp,GS:20\%,$n=102$ & 23,488 & 6,057 & 2,806 & 12,456 & 2,162 & 0.737 & 0.816 & 0.788 & 0.684 & 0.545 & 4.023 & 0.322  \\ 
weighted, exp/exp,GS:33\%,$n=171$ & 23,488 & 6,182 & 2,852 & 12,409 & 2,036 & 0.752 & 0.813 & 0.792 & 0.684 & 0.554 & 4.029 & 0.305  \\ 
weighted, exp/exp,GS:50\%,$n=257$ & 23,488 & 6,262 & 2,898 & 12,364 & 1,957 & 0.762 & 0.810 & 0.793 & 0.684 & 0.559 & 4.013 & 0.294  \\\hline
	
\end{tabular}
}
\end{table*}

\begin{table}[h]
\caption{\label{table:wilcox} p-values obtained using a pairwise Wilcoxon test comparing the classification times between groups. }
\centering
\begin{tabular}{rrrr}
  \hline
 & beginner & competent & experienced \\ \hline
	competent   & 1.0000 &  &  \\ 
  experienced & 0.5932 & 0.2955 &  \\ 
  expert      & {\bf 0.0037} & {\bf 0.0009} & 0.0781 \\ \hline
\end{tabular}
\end{table}

\clearpage
\section{Supplementary Materials}

\clearpage

% big table
\newgeometry{inner=1cm,outer=1cm, top=1cm, bottom=1cm}

\begin{landscape}

\subsection*{Summary of Bayesian methods in Citizen Science}
%\rowcolors{1}{white}{gray(x11gray)} %lightgray

\begin{longtable}{p{2cm}p{3cm}p{5cm}p{2.5cm}p{4.85cm}p{3.5cm}} 

\caption{Summary of publications. 
%The method refers to a classification from Table \ref{table:symbols}. 
%In the column description, we succinctly describe the Bayesian technique used. The software used for the computations and package (if mentioned) are also included. The column data briefly describes where the data is coming from and the period of time under analysis.
} \\\hline

\endfirsthead

\caption[]{Summary of publications included in the review. 
} \\\hline
\endhead
% Data: Species, Location. numb citizens. numb items. 
% in space  , in time
% ecology
% bias correction

Author(s) & Method & Description & Software / package  & Data & Platform/ Project \\ \hline %  & Application

\endhead
Author(s) & Method & Description & Software / package & Data & Platform\\ \hline % 

\citet{WOS:000786312200001}      & Species distribution model & Assessing and predicting Japanese beetle species distributions to identify potential invasion locations & \textsf{R}(\textsf{rinat}, \textsf{adehabitatHR}, \textsf{mclust}, \textsf{envirem}) &  International occurrence records for the Japanese beetle (2010-2020) & iNaturalist \\ \hline
% P-ID/YR: DELLA ROCCA, 2022
% P-FOCUS: Assessing and predicting Japanese beetle species distributions to identify potential invasion locations.
% P--DATA: International occurrence records for the Japanese beetle (2010-2020).
% P-MODEL: Species distribution model, spatio-temporal (5km grid globally by year from 2010-2019).
% M-EXTRA: Unspecified priors. Bias mentioned, not addressed.
% P--SOFT: R (rinat, adehabitatHR, mclust, envirem), R-INLA.
% P--PLAT: iNaturalist.

\citet{WOS:000779469800001}      & Connectivity analysis (Bayes' rule) & Methodology for estimating patterns of migratory connectivity for songbirds using relative abundance models & \textsf{R} (\textsf{adehabitat}, \textsf{NbClust}, \textsf{cluster}) &  Relative abundance estimates data for wood thrush and Wilson's warbler songbirds (2016-2017, international) & eBird \\ \hline
% P-ID/YR: VINCENT, 2022
% P-FOCUS: Methodology for estimating patterns of migratory connectivity for songbirds using relative abundance models.
% P--DATA: Relative abundance estimates data for wood thrush and Wilson's warbler songbirds (2016-2017, international).
% P-MODEL: Connectivity analysis (Bayes' rule), no space or time components.
% M-EXTRA: No priors. No mention of biases.
% P--SOFT: R (adehabitat, NbClust, cluster).
% P--PLAT: eBird.

\citet{WOS:000740375100001}      & Bayesian hierarchical model & Using float-plane flight pathway data to inform early detection of aquatic invasive species & \textsf{R} (\textsf{rstanarm}) & Flight destination data (2015-2016, Alaska, North America) &  \\ \hline
% P-ID/YR: SCHWOERER, 2022
% P-FOCUS: Using float-plane flight pathway data to inform early detection of aquatic invasive species.
% P--DATA: Flight destination data (2015-2016, Alaska, North America).
% P-MODEL: Bayesian hierarchical model, no space or time components.
% M-EXTRA: Normal priors (weakly informative, uninformative)
% P--SOFT: R (rstanarm), R-Stan.
% P--PLAT: None.

\citet{santos2021} &  Spatial item response model for image classification   &	Produce estimates of the participants abilities adjusting for the difficulty of the image.
Implemented a divide-and-conquer approach to fit the model to big datasets.
&	\textsf{R} (\textsf{rtan}, \textsf{mirt})	 &  Camera traps images from the Serengeti, Africa classified by volunteers. Big datasets of 10.8 million classifications produced by 28,000 users  &   \url{https://www.zooniverse.org/projects/zooniverse/snapshot-serengeti} \\ \hline

%\citet{santos2021}      & Bayesian item response model & Estimation of user accuracy in image classifications. Accounts for spatial dependence & R (Stan, mirt) & & Species identification data for various species (2015, Serengeti, Tanzania) & Zooniverse \\ \hline
% P-ID/YR: SANTOS-FERNANDEZ, 2021
% P-FOCUS: Estimation of user accuracy in image classifications.
% P--DATA: Species identification data for various species (2015, Serengeti, Tanzania).
% P-MODEL: Item response model, spatial (site in Tanzania).
% M-EXTRA: Various priors. Biases mentioned, not addressed.
% P--SOFT: R (mirt, Stan), R-Stan.
% P--PLAT: Zooniverse.

\citet{WOS:000663754100002} & Independent Bayesian Classification Combination (IBCC) model & Bayesian methods for the estimation of iconic taxa classification accuracy & Unknown  & Observation and identification data for 13 species of iconic taxa (2012-2017, New Zealand) & iNaturalist NZ \\ \hline
% P-ID/YR: MUGFORD, 2021
% P-FOCUS: Bayesian methods for the estimation of iconic taxa classification accuracy.
% P--DATA: Observation and identification data for 13 species of iconic taxa (2012-2017, New Zealand).
% P-MODEL: Independent Bayesian Classification Combination (IBCC) model. No space or time components.
% M-EXTRA: Beta priors (informed by existing user accuracy). No mention of biases.
% P--SOFT: Unknown.
% P--PLAT: iNaturalist NZ.

\citet{sheard2021long} & Bayesian occupancy model   & Identifying temporal trends in ant species occupancy in Denmark  & \textsf{R} (\textsf{sparta}, \textsf{R2jags})  & Detection data for 29 species of ant (1903-2019, Denmark)    & The Ant Hunt, EuroAnts, museum records, and personal observations \\ \hline
% P-ID/YR: SHEARD, 2021
% P-FOCUS: Identifying temporal trends in ant species occupancy in Denmark.
% P--DATA: Detection data for 29 species of ant (1903-2019, Denmark).
% P-MODEL: Bayesian occupancy model, spatio-temporal (10km grid cells in Denmark per decade from 1900-2019).
% M-EXTRA: Normal priors (vague, uninformative). No mention of biases.
% P--SOFT: R (sparta, R2jags), R-JAGS.
% P--PLAT: The Ant Hunt, EuroAnts, museum records, and personal observations.

\citet{WOS:000719372700005}      & Bayesian occupancy model & Estimates of long-term occupancy trends for terrestrial mammal species  & \textsf{R} (\textsf{R2jags}, \textsf{sparta}) & Biological records of terrestrial mammals (1970-2017, United Kingdom) & The Mammal Society (data set) \\ \hline
% P-ID/YR: COOMBER, 2021
% P-FOCUS: Estimates of long-term occupancy trends for terrestrial mammal species.
% P--DATA: Biological records of terrestrial mammals (1970-2017, United Kingdom).
% P-MODEL: Bayesian occupancy model, spatio-temporal (site in United Kingdom per year from 1970-2016).
% M-EXTRA: Normal priors (widely dispersed). No mention of biases.
% P--SOFT: R (R2jags, sparta), R-JAGS.
% P--PLAT: The Mammal Society (data set).

%\citet{WOS:000691436000001}      & Species distribution models  &     & SOFTWARE/PACKAGE  & virtual species  & simulated presence-absence and presence-only data & PLATFORM/PROJECT \\ \hline
% P-ID/YR: CHEVALIER, 2021 (not included: only data used is simulated)
% P-FOCUS:
% P--DATA:
% P-MODEL:
% M-EXTRA:
% P--SOFT:
% P--PLAT:

\citet{WOS:000686044200001}      & Species distribution model  & Integration of spatial autocorrelation to estimate sampling effort and model species densities & Unknown  & Sighting event data for fur seals and Stellar sea lions (1958-2016, Bering Sea and Gulf of Alaska) & Platforms of Opportunity (data set) \\ \hline
% P-ID/YR: VER HOEF, 2021
% P-FOCUS: Integration of spatial autocorrelation to estimate sampling effort and model species densities.
% P--DATA: Sighting event data for fur seals and Stellar sea lions (1958-2016, Bering Sea and Gulf of Alaska).
% P-MODEL: Species distribution model, spatial (site in Bering Sea or Gulf of Alaska).
% M-EXTRA: Unspecified priors (flat). Modelled observer bias, estimated spatial density to address sampling bias.
% P--SOFT: Unknown
% P--PLAT: Platforms of Opportunity (data set).

\citet{WOS:000627494200001}       & Bayesian hierarchical models & Bird population estimated using combined structured and unstructured citizen science data sets & \textsf{R} (\textsf{rtrim}, \textsf{greta}) & Observation data for a range of birds (2012-2019, Germany) & Ornitho, eBird, Observation \\ \hline
% P-ID/YR: HERTZOG, 2021
% P-FOCUS: Bird population estimated using combined structured and unstructured citizen science data sets.
% P--DATA: Observation data for a range of birds (2012-2019, Germany).
% P-MODEL: Bayesian hierarchical models, spatio-temporal (site in Germany per year from 2012-2019).
% M-EXTRA: Unspecified priors. Biases mentioned, not addressed.
% P--SOFT: R (rtrim, greta), R-TRIM
% P--PLAT: Ornitho, eBird, Observation.

\citet{WOS:000668586900001}      & Bayesian occupancy model    & Incorporating unstructured herbarium and citizen science data into an occupancy modeling framework, with bias correction techniques   & \textsf{R} (\textsf{MCMCvis}), \textsf{openRefine}  &    Occurrence data for Anacardiaceae and Tracheophytes (2019, Florida, North America)    & Global Biodiversity Information Facility (GBIF), iNaturalist, naturegucker, Atlas of Living Australia's Questagame \\ \hline
% P-ID/YR: ERICKSON, 2021
% P-FOCUS: Incorporating unstructured herbarium and citizen science data into an occupancy modeling framework, with bias correction techniques.
% P--DATA: Occurrence data for Anacardiaceae and Tracheophytes (2019, Florida, North America).
% P-MODEL: Bayesian occupancy model, spatial (county in Florida).
% M-EXTRA: Beta priors (flat). Bias addressed by combining data-level and model-level bias correction techniques.
% P--SOFT: openRefine, R (MCMCvis)
% P--PLAT: Global Biodiversity Information Facility (GBIF), iNaturalist, naturegucker, Atlas of Living Australia's Questagame.

%\citet{vitense2018uncovering}      & METHOD    & DESCRIPTION   & SOFTWARE/PACKAGE  & APPLICATION   & DATA    & PLATFORM/PROJECT \\ \hline
% P-ID/YR: VITENSE, 2018 (not included: not citizen science)
% P-FOCUS: Classifying lake states, identifying critical total phosphorus thresholds, and estimating steady state relationships between total phosphorus thresholds and chlorophyll.
% P--DATA: 
% P-MODEL: Bayesian latent variable regression (BLR).
% M-EXTRA:
% P--SOFT:
% P--PLAT: Minnesota Department of Natural Resources (data set).

\citet{rodhouse2021audible} & Bayesian multi-season occupancy model & Updated species distribution model for the rare spotted bat using new survey data & \textsf{R} (\textsf{R2jags}) & Acoustic capture encounter records for 14 bat species (2003-2010, 2019-2020), presence-only encounter records for E. maculatum (1976-2018, Oregon, North America) & NABat \\ \hline
% P-ID/YR: RODHOUSE, 2021
% P-FOCUS: Updated species distribution model for the rare spotted bat using new survey data.
% P--DATA: Acoustic capture encounter records for 14 bat species (2003-2010, 2019-2020), presence-only encounter records for E. maculatum (1976-2018, Oregon, North America).
% P-MODEL: Bayesian multi-season occupancy model, spatio-temporal (grid in Oregon by season from 1976-2020).
% M-EXTRA: Unspecified (informative, vague). No mention of biases.
% P--SOFT: R (R2jags), R-JAGS.
% P--PLAT: NABat.

\citet{WOS:000738781900001}      & Bayesian kriging & Modelling the spatial distribution of crowd-sourced radiation measurements surrounding Fukushima & \textsf{R} (\textsf{stats}, \textsf{raster}, \textsf{SpatialEco}, \textsf{hexbin}, \textsf{fields}, \textsf{lattice}) & Coordinate-based radiation levels (2011-2015, Fukushima, Japan) & Safecast, United States Department of Energy (data set) \\ \hline
% P-ID/YR: HULTQUIST, 2021
% P-FOCUS: Modelling the spatial distribution of crowd-sourced radiation measurements surrounding Fukushima.
% P--DATA: Coordinate-based radiation levels (2011-2015, Fukushima, Japan).
% P-MODEL: Bayesian kriging, spatial (point data in Fukushima).
% M-EXTRA: No priors. No mention of biases.
% P--SOFT: R (stats, raster, SpatialEco, hexbin, fields, lattice).
% P--PLAT: Safecast, United States Department of Energy (data set).

\citet{WOS:000667245100032}      & Bayesian maximum entropy & Estimating spatio-temporal distribution of depth to groundwater in remote Columbia & Unknown & Well water depth measurements (2008-2009, Bajo Cauca Region, Columbia, South America) &  \\ \hline
% P-ID/YR: GOMEZ, 2021
% P-FOCUS: Estimating spatio-temporal distribution of depth to groundwater in remote Columbia.
% P--DATA: Well water depth measurements (2008-2009, Bajo Cauca Region, Columbia, South America).
% P-MODEL: Bayesian maximum entropy, spatio-temporal (point data in the Bajo Cauca Region by week from 2008-2009).
% M-EXTRA: Normal priors. Sampling bias addressed by implementing models on data subsets (using a space-time soft data approach).
% P--SOFT: Unknown.
% P--PLAT: None.

\citet{regnier2021age}      & Von Bertalanffy growth function (VBGF) & Estimating population of the critically endangered skate using photo-identification of trophy pictures and tag-recapture data & \textsf{R} (\textsf{rjags}) &  Capture-mark-recapture (CMR) data for skates (2016-2018, Scotland) & NatureScot, Scottish Association for Marine Science (data set) \\ \hline
% P-ID/YR: REGNIER, 2021
% P-FOCUS: Estimating population of the critically endangered skate using photo-identification of trophy pictures and tag-recapture data.
% P--DATA: Capture-mark-recapture (CMR) data for skates (2016-2018, Scotland).
% P-MODEL: Von Bertalanffy growth function (VBGF), no space or time components.
% M-EXTRA: Normal and t-distribution priors (vague). No mention of biases.
% P--SOFT: R (rjags), R-JAGS.
% P--PLAT: NatureScot, Scottish Association for Marine Science (data set).

\citet{WOS:000647892500001}      & Bayesian assignments    & Measuring migratory connectivity for five songbird species  & ArcGIS 10.1 (Spatial Analyst), R (\textsf{stats}, \textsf{wBoot}, \textsf{ade4})  & Feather sample data for five songbird species (1995-2010, Caribbean Basin) & American Breeding Bird Suvery (BBS) \\ \hline
% P-ID/YR: STUDDS, 2021
% P-FOCUS: Measuring migratory connectivity for five songbird species.
% P--DATA: Feather sample data for five songbird species (1995-2010, Caribbean Basin).
% P-MODEL: Bayesian assignments, spatial (site in Caribbean Basin).
% M-EXTRA: No mention of priors. No mention of biases.
% P--SOFT: Spatial Analyst, R (stats, wBoot, ade4).
% P--PLAT: American Breeding Bird Suvery (BBS).

\citet{eisma2020bayesian} & Measurement error model & Developed a graphical inference model for assessing the accuracy of citizen scientists to report depth of rainfall. The verification was made using a photo of the rain gauge. They reported estimates from several error types.   &  \textsf{Infer.NET}     & 153 citizen scientists equipped with a rain gauge and smart phones &  Nepal rainfall observations program.  \\ \hline %
%There exist
% error model. respsnse variable is a continous measurement, hence they used a Measurement error model 
% reported the depth of rainfall in millimeters
% verification using a photo of the rain gauge.
% recorded several demographic covariates (age, education, ocupation, experience, motivation (paid or volunteer), etc)  

\citet{peterson2020monitoring} & Spatio-temporal weighted beta regression model  	&	 Models the proportion of hard corals as a function of ecological disturbances and management decisions. It integrate data from professional and non-professional data sources &	\textsf{R}, \textsf{INLA}	 	 &  218 citizen-contributed images from Reef Check Australia classified by 12 citizen scientists  & Virtual Reef Diver \url{https://www.virtualreef.org.au}\\ \hline

\citet{allen2020counting} & Zero-inflated Poisson (ZIP) models in an a-priori modeling framework (spatial)		 & Estimate leopard population densities in the Serengeti		 &  \textsf{R} (\textsf{pscl}), \textsf{ArcGIS}	 &  Camera traps placed in randomly distributed 5kmsq grid, one year data (wet/dry season) &  Snapshot Serengeti Project \url{https://www.zooniverse.org/projects/zooniverse/snapshot-serengeti}\\ \hline
% estimate leopard population densities in the serengeti 
% determine potential biological drivers for abundance and distribution 
% "Bayesian framework"
% largest citizen science project - Snapshot Serengeti Project - camera traps (Swanson et al, 2015) 
% spatially-explicit capture-recapture models for unmarked populations (2012 project/paper - look it up)
% zero-inflated Poisson (ZIP) models in an a-priori modeling framework 
% 5km^2 grid, 250m from the centre of each grid - cameras 
% abundance = events per 100 trap nights
% ArcGIS for habitat variables - Tanzania sentinel spatial data https:// geoportal.rcmrd.org/layers/servir%3Atanzania_sentinel2_lulc2016
% hydrology/rivers from LANDSAT https://www.fao.org/geonetwork/srv/en/main.home
% used other studies of similar nearby systems to estimate parameters

\citet{crawford2020estimating} 		& Cormack-Jolly-Seber model 	& Estimating population persistence for at-risk species using citizen science data. 		 & \textsf{R} (\textsf{R2jags}), \textsf{ArcGIS}		 &   HerpMapper observations of multiple snake species. Hognose snake was at-risk species that only accounted for 2.5\% of observations. & HerpMapper \url{https://www.herpmapper.org/} \\ \hline
%Estimating population persistence for at-risk species using citizen science data - hognose snake
% software : R, ArcGIS
% data : HerpMapper - validated by professional herpetologists - oppportunisitc sighitings, observations from systematic stufies from trained researchers 
% not just hognose snake but other snakes from HerpMapper in Florida - these are the non-target species - indicates when an area was  searched but no hognose was found 
% filtered observers to only 1 obs per day (some would submit multiple records on the same date) 
% Cormack-Jolly-Seber model - mark recapture! annual population persistence probability and detection probability
% what is a Bernoulli trial? 
% hognose only 2.5% of herp records 
% model lacks a recolonisation mechanism (typical in occupancy models) - preliminary models showed poor convergence so didnt do it - however this decision could lead to underestimates of persistence... assumed colonisation rates are neglible 
% hognose has a relatively low dispersal distance and is difficult to detect - so the post 1950 records might be because of difficulty detecting instead of population decline 
% software : R (R2jags)
% used a projection model to predict persistence through 2080 - did some interesting stressor scenarios and environmental prediction for urbanisation etc - multiloop stochastic simulation model - used predict.glm for habitat suitability prediction 
% doesnt really discuss method/math - very application based paper
% good for when data is limited but conservation decisions have to be made - looking at extinction risk 

\citet{kagawa2020citizen} 			& Hierarchical Bayesian approach  	& Using social media to monitor the symbiosis between marine gastropod and epibiotic alga. 	 & \textsf{R} (\textsf{fetchR}, \textsf{rstan}), \textsf{QGIS}		 &  Social median submissions (hastags, Google forms) along Japan and Korea coastlines  & Suagi \& Kaigoromo National Survey \url{https://sites.google.com/view/sugai/}\\ \hline
%Citizen science via social media revealed conditions of symbiosis between a marine gastropod and an epibiotic alga
% alga grows on shell of gastropod - symbiotic relationship 
% Japan and Korea coastline - social media (citizen science) (https ://sites .googl e.com/view/sugai /)
% Hierarchical Bayesian model 
% proportion of shell increased as the substrate size of gastropod increased 
% MCMC sim for posterior distribution  
% no discussion of the math/method - discussion before the method - odd paper format 
% beta distribution 
% software : R (fetchR, R-Stan), QGIS 
% Spatio-temporal 

\citet{lavariega2020community} 		& Spatially explicit capture-recapture. Single-season occupancy-detection model with MLE framework 	& Framework for community based monitoring of Jaguar in Mexico. Spatially explicit capture–recapture models estimated a density using Bayesian methods. 		 &	\textsf{R} (\textsf{SECR}, \textsf{SPACECAP}, \textsf{unmarked}, \textsf{AICcmodavg}), \textsf{QGIS}	 &  Camera-traps in 5 local communities with 8 Jaguar sightings   &  \\ \hline
%Community-Based Monitoring of Jaguar (Panthera onca) in the Chinantla Region, Mexico
% Camera trap survey - 8 jaguars! Spatially explicit capture–recapture models estimated a density using Bayesian methods 
% Jaguar 'captures' locations and habitat used influenced by vegetation and capture rate of prey 
% the community-based monitoring program increased the communication and interaction of local community - increased participants knowledge 
% this paper is the framework with hopes of making a large scale monitoring program across Mexico 
% 5 local communities - given training by expert for camera traps 
%Jaguar density was estimated using spatially explicit capture-recapture (SCR) modeling methods
%SCR models are less biased by the edge effects and they do not require an ad hoc estimation of the effective area surveyed, as is necessary with the traditional capture–recapture models
% two models 1) distribution of animals 2) observation process 
% two approaches to estimate derived parameters: 1) a likelihood-based method in the SECR package 2) a Bayesian inference method in the SPACECAP package - both fit a state-space model and observation model with imperfect spatial detection 
% Software : R (SECR, SPACECAP, unmarked, AICcmodavg)
% doesnt discuss the method and math that much - mainly talks about the functionality of the R packages - method section is dense and bad poorly formatted 
% spatially stuff done in QGIS 
% model selection with AIC, chisqd - occupancy-detection models were fit with 'unmarked' package

\citet{louvrier2020mechanistic} 	& SDM, Bayesian, PDE for dispersion 	& Mechanistic–statistical species distribution model to explain and forecast wolf (Canis lupus) colonization in South-Eastern France.		 & \textsf{JAGS}, \textsf{R} (\textsf{mecastat}), \textsf{OpenBUGS}		 &   Professional and non-professional wolf observers collected presence signs all year round (2007-2015) &  \\ \hline
%A mechanistic–statistical species distribution model to explain and forecast wolf (Canis lupus) colonization in South-Eastern France
% standard SDMs dont incorporate dynamic processes like dispersal - potential for bias in inference about species distributions 
% ecological diffusion - incorporates spatio-temporal processes in ecological models 
% recolonisiation of wolves in eastern france - naturally from dispersion since the 1990s 
% PDEs for modeling species diffusion and growth - developed a mechanistic-statistical spatio-temporal model to account for ecological diffusion, logistic growth and imperfect species detection 
% growth rate influenced by forest cover
% diffusion influenced by human density 
% species detectability increased with increased survey effort 
% state-space modeling approach 1) observation process that handles the stochasticity in the detections and non-detections conditional on 2) latent state process which is described by mechanistic model 
% software: JAGS (mecastat package) and OpenBUGS
% actually discusses the model and math 
% data : professional and non-professional wold observers 
% Presence signs went through a standardized control process to prevent misidentification
% spatial - 10km by 10km resolution grid but estimated the parameters based on a 3 by 3km grid - 10km low error compared to 3km 

\citet{maguire2020citizen} 			& Dynamic occupancy model (DOM)  	& Temperature-driven declines in riverine sentinel invertebrates. Funding dependent surveys with sporadic sampling (2002-2018). Missing values modelled with spatial stream networks. 		 &	\textsf{\textsf{R} (\textsf{h2o}, \textsf{SSN})}	 &  	Stonefly surveys lead by expert with citizen scientists (2002-2018)   & \\ \hline
% Stonefly in Detroit, over 17 years (2002-2018), citizen science data
% surveys based on sporadic funds - inconsistent, missing values 
% missing values modelled - not every site was sampled every year 
% spatio temporal 
% River Rouge in urban Detroit 
% water temperature and landuse data to model missing values 
%dynamic occupancy model (DOM) - occurrence, extinction, colonization, and detection probabilities for winter stoneflies
% stoneflies most impacted by water quality BUT no water quality data was collected - proxy is landuse 
% Machine learning to estimate the 25 stonefly sites
% random forest model - 198 landuse variables 
% software: h2o, R
% used SSN (Erin's package!) - year as random effect, temperature as fixed effect - the pres/absence extracted from SSN and used in DOM
% SSN to fill in missing values, then used DOM (Bayesian)
% Bayesian state-space model of daily mean stream temperature
% the citizen scientists are highly trained through a quality assurance project plan - there is a trained team leader that oversees each group of citizen scientists 

\citet{sicacha2020accounting} 		& Bayesian Occupancy	& Accounting for spatial varying sampling effort due to accessibility of Citizen Science data. Inference about the important of terrain ruggedness index and solar radiation on Moose. 		 &	\textsf{INLA}	 &   472 observations of Moose in Norway   & GBIF \url{https://www.gbif.org/} \\ \hline
% Moose in Norway
%Accounting for spatial varying sampling effort due to accessibility in Citizen Science data
% aim: make an inference about the importance of terrain ruggedness index and solar radiation on Moose
% discusses strong spatial bias in survey effort - overrepresentation of occurances close to roads (use distance to road as proxy for accessibility)
%Bayesian Log-Gaussian Cox Process specification for occurrence
% data : GBIF (https://gbif.org) - citizen science platform 
% spatial - 1km by 1km resolution, monthly solar radiation average (temporal), terrain ruggedness 
% 472 observation points 
% look up: inhomogeneous Poisson Process (NHPP)
% section : Naive model, variation in sampling effort (VSE) model, extended variation in sampling effort model (EVSE)
% software : INLA 
% model assessment, used DIC, WAIC and LPML

\citet{zhu2020migratory} 			& Bayesian SDM 	& Migratory connectivity of Swan Geese from SDMs, feather stable isotope assignment and satellite tracking		 & \textsf{MAXENT}, \textsf{R} (\textsf{maptools},\textsf{raster})		& Collected 96 feathers from capturing swan geese in breeding and wintering grounds (2010-2017)  & eBird \url{https://ebird.org/home}, GBIF \url{https://www.gbif.org/} \\ \hline

\citet{santos2020correcting} &  Bayesian hierarchical model, Spatial beta regression 	& Corrects misclassification errors in the annotation of images.
Adjust estimates of the proportion hard coral based on citizens abilities or performance measures.
It accounts for spatial variation. 		 &	\textsf{R} (\textsf{stats}, \textsf{rstan}), \textsf{Stan}	 &  	514 images from the Great Barrier Reef classified by 223 citizens. Amazon Mechanical Turk classifications.  & Virtual Reef Diver \url{https://www.virtualreef.org.au} \\ \hline
%Correcting misclassification errors in crowdsourced ecological data:a Bayesian perspective.
% learn latent variables of interest to account for paticipants abilities - eg coral cover 
% participant's abilities are expressed in terms with sensitivity and specificity of a correctly classified set of points 
% spatial model 
% Stan, R (stats package)
% GLM with a beta distributed response 
% bias approaches 1) ignore measuremnt error 2) weighted linear regression 3) imperfect classification taking into consideration user' sensitivity and specifity (true positive, true negative) 
% spatially dependent misclassification error - paper's method 
% The great barrier reef, https://www.virtualreef.org.au
% 1585 images, 2008-2017 
% covariates - DHW, not take marine reserve, shelf position, cyclone effect 
% Voronoi polygons for spatial domain 
% Amazon Mechanical Turk used for data collection 
% help file for citizens and pass qualification test 
% methods actually discusses the math and graph of weighted model
% Method paper - focused on bias correction  

%\citet{santos2019error} &  Spatial beta regression  &	Adjust estimates of the proportion hard coral based on citizens abilities or performance measures 	 &
%\textsf{R, Stan	} &  spatial	 &  514 images classified by 223 citizens   &  Virtual Reef Diver \url{https://www.virtualreef.org.au} \\ \hline

\citet{WOS:000604981300010}      & Bayesian hierarchical model    & Population trajectory estimation for North American birds   & \textsf{R} (\textsf{brms}, \textsf{rstan}) &  Identification data for 28 species of songbirds and raptors (1928-2016, North America)   & eBird \\ \hline
% P-ID/YR: WALKER, 2020
% P-FOCUS: Population trajectory estimation for North American birds.
% P--DATA: Identification data for 28 species of songbirds and raptors (1928-2016, North America).
% P-MODEL: Bayesian hierarchical model, spatio-temporal (40km grid cells in North America per season from 1928-2016).
% M-EXTRA: Unspecified priors (improper, flat). Biases mentioned, not addressed.
% P--SOFT: R (brms, Stan), R-Stan.
% P--PLAT: eBird.

\citet{benshemesh2020citizen}   & Bayesian hierarchical model    & Estimations of trends and drivers of malleefowl breeding activity   & \textsf{R} (\textsf{R2OpenBUGS}, \textsf{coda})  & Malleefowl mound activity data (1989-2017, VIC SA, WA, and NSW, Australia) & National Malleefowl Monitoring Database (NMMD) \\ \hline
% P-ID/YR: BENSHEMESH, 2020
% P-FOCUS: Estimations of trends and drivers of malleefowl breeding activity.
% P--DATA: Malleefowl mound activity data (1989-2017, VIC SA, WA, and NSW, Australia).
% P-MODEL: Bayesian hierarchical model, spatio-temporal (site in Australia per year from 1989-2017).
% M-EXTRA: Normal priors (vague). Biases mentioned, not addressed.
% P--SOFT: R (R2OpenBUGS, coda).
% P--PLAT: National Malleefowl Monitoring Database (NMMD).

\citet{WOS:000514248900010}      & Bayesian generalised linear model & Using multiple online data platforms to determine which parameters impact the location of bird sightings & \textsf{R} (\textsf{sp}, \textsf{rjags}, \textsf{AER}, \textsf{spdep}) & Bird observation data (2015-2017, Chicago, North America) & iNaturalist, eBird, Flickr \\ \hline
% P-ID/YR: LOPEZ, 2020
% P-FOCUS: Using multiple online data platforms to determine which parameters impact the location of bird sightings.
% P--DATA: Bird observation data (2015-2017, Chicago, North America).
% P-MODEL: Bayesian generalised linear model, spatial (by community area in Chicago).
% M-EXTRA: Normal priors (uninformative). Biased mentioned, not addressed.
% P--SOFT: R (sp, rjags, AER, spdep), R-JAGS.
% P--PLAT: iNaturalist, eBird, Flickr.

\citet{ryan2019global} & Approximate Bayesian Computation (ABC) based on Random Forests & Modelling the genetic structure of the population  & \textsf{fastSTRUCTURE} (written in \textsf{Python}2.x), \textsf{VCFtools}, \textsf{R} (\textsf{adegenet}, \textsf{poppr}, \textsf{ape}) &  Small cabbage white butterfly (Pieris rapae). 32 countries. 150 volunteer scientists and citizens contributing a substantial proportion of the data.
\url{https://github.com/citscisean/PierisrapaeInvasionHistory}  & Pieris Project (\url{http://www.pierisproject.org/})\\ \hline % 

\citet{humphreys2019seasonal} & Bayesian regression with a hierarchical framework. Logistic and Poisson regression models in space and time & Estimation of the occurrence and abundance. Effect of predictors such as precipitation, continentality and wetness on dabbling ducks population distributions & \textsf{R} (\textsf{r‐INLA}, \textsf{raster}) & ~ 1 million records from ten ducks species in the United States from 2015-2017 & eBird project (\url{https ://ebird.org/home})\\ \hline % 

% Spatially explicit mode
% Species distribution model
% citizens provide the images i think 
% covariates temperature and precipitation data,
%Gaussian random fields to model spatial random effects. first order random walk mode

\citet{croft2019modeling} & Hierarchical Bayesian model &  Species distribution models for presence‐only data in six deer species. Accounts for spatial autocorrelation. & \textsf{R} (\textsf{hSDM}) & Great Britain (GB).  National Biodiversity Network (NBN) Atlas. Data from 2012-2016.  & \\ \hline % 

\citet{cumming2019point} & Bayesian multispecies occupancy model 	& Models bird point counts across South African National Parks. It accounts for imperfect detection 		 &	\textsf{R} (\textsf{jagsUI}, \textsf{AHMbook})	 &  	  & \\ \hline
% South African birds - point counts 
% line transects on 302 routes in 19 South African National Parks
% Bayesian multispecies occupancy model - accounts for imperfect detection
% geo referenced observations 
% walking/driving routes 
% four measures of species richness. line transects - absolute number and effort corrected number of species recorded. point counts - absolute number and estimated number of species by occupancy model based on detection/non-detection data 
% effort corrected values were standardised by correcting for time spent and area covered 
% "To correct for observing effort, we therefore (1) regressed observed species richness on observing time by route; and (2) extracted the residuals to give a new variable, time-corrected transect richness"
% measure mean detection and occupancy probabilities for each species and estimate species richness
% substituted space for time
% JAGS, R (jagsUI, AHMbook) 
% species richness was severely underestimated when survey replicates were unavailable and imperfect detection was not taken into account
% Hierarchical? first mentioned in discussion
% obtained greater accuracy with point counts rather than line transects

\citet{girardello2019gaps} & Hierarchical Bayesian and spatial linear models 	&	Discusses observational and geographic bias 	  &	INLA \textsf{R} (\textsf{vegan}, \textsf{fossil})	 &    & Biodiversity Information Facility (GBIF) \url{https://www.gbif.org/occurrence/download/0028634-181108115102211}\\ \hline
% gaps in butterfly data 
% 19 million species occurrence records - database 
% discusses observational bias and geographic bias 
% butterflies were picked because of good records as they are pretty, popular and easy to spot 
% READ:: (Gustafsson et al., 2015; Kocher and Williams, 2000)
% assessed inventory completeness with smoother species accumulation curves (SACs) 
% GLMs: assess country trends. fit to total number of records per country (aggregated to yearly level). Poisson, log link 
% spatial linear models fit to response variable 
% Bayes - INLA
% spatially correlated random intercept (GMRF) 
% showed that spatial bias exists - locations with appeal to researchers, accessibility etc 
% cool incompleteness map plot 

\citet{lyon2019increased} & Bayesian hierarchical model and state-space Cormack-Jolly-Seber model	& Uses a state-space Cormack-Jolly-Seber approach to model  population size of Murray cod and gold perch.	 &	\textsf{JAGS}, \textsf{R}	 &   Captured 7,312 Murray cod and 3,743 golden perch (2007-2013) from electrofishing census and tagging on 3,829 Murray cod and 3,316 golden perch with PIT and external tags and 689 Murray cod and 466 golden perch with radio tags. Anglers reported 1,338 tagged Murray cod and 275 golden perch.  & \\ \hline

\citet{outhwaite2019annual} & Bayesian occupancy model & Occupancy and species trends for bryophytes, lichens, and invertebrates in the UK. &		 &   5,293 UK bryophytes, lichens, and invertebrates (1970-2015)  & \\ \hline
% occupancy and species trends for bryophytes, lichens, and invertebrates
% Bayesian occupancy model 
% a lot of 'priority' species have been monitored closely, but this paper wants to use this data to look at the lesser priority species (lack of abundance data) 
% biological records - presence only - collected by volunteers and collated by recording schemes and societies 
% at least 4 forms of bias : the uneven detectability of species across space and time, uneven sampling effort per visit, uneven spatial coverage and uneven recording intensity over time
% bias creates challenge for estimating temporal change 
% extended a previous Bayesian occupancy modelling framework, to increase the precision of occupancy estimates via the use of a random walk prior on the year effect of the state model
% at fine scale 1km by 1km resolution 
% data collated from 29 recording schemes and societies, and from Biological Records Centre, Wallingforth and iRecord \url{https://www.brc.ac.uk/irecord/}, NBN Atlas https://nbnatlas.org/
% BRC scheme pages (https://www.brc.ac.uk/recording-schemes)

% cite & method & description & software & application & data & platform 
\citet{viljugrein2019method} 		& Hierarchical change-in-ratio model, Stochastic scenario tree &  Models relationship between diagnostic test sensitivity, sample quality and disease progression. Studies chronic wasting disease (CWD) infections in European reindeer populations. 		 &	\textsf{R} (\textsf{fitdistrplus}, \textsf{R2Jags})	   	 & Citizen (hunters) collected heads for brain tissue extraction by experts  &  \\ \hline
% chronic wasting disease (CWD) in cervids (deer) 
% % stochastic scenario tree model of diagnostic sensitivity 
% apply the diagnostic sensitivity in a Bayesian framework - enables estimation of age-class-specific true prevalence 
% population of European reindeer 
% collected brain tissue only or both brain and lymph node samples
% collected by both trained professionals and hunters (our citizen scientists)
% simulation model (scenario tree) to estimate the test sensitivity of CWD because of the mixed samples (complex error structure) 
% hierarchical change-in-ratio model - based on aerial minimum counts in winter, calving surveys in summer, harvest data and demographic structure counts in fall 
% 
%

\citet{patten2019breeding} 			& Bayesian occupancy model 	& Breeding thresholds in opportunistic Odonata records		 &	\textsf{R} (\textsf{unmarked}, \textsf{rjags}, \textsf{segmented}), \textsf{ArcMap}	 &   Citizen scientist image accompanied submissions from two online portals (2013 - 2017) & Odonata Central \url{https://www.odonatacentral.org/}, iNaturalist \url{https://www.inaturalist.org/}\\ \hline
%Breeding thresholds in opportunistic Odonata records
% Odonata records - dragonflies 
%logistic curves from occupancy models of observations of tenerals (young adults) - subjected curves to Bayesian two-segment piecewise regression 
% >20,000 geocoded adult odonate records obtained from 2013 to 2017 across Oklahoma, USA
% Oklahoma Odonata Project (OOP) database in Oklahoma Natural Heritage Inventory at the Oklahoma Biological Survey
% limited it purposely to 5 recent years - these records have more details about age, sex, behaviour 
% citizen scientists submitted data to OOP or two online portals Odonata Central (https://www.odonatacentral.org/) and iNaturalist (https://www.inaturalist.org/) - images to accompany records 
% each record is vetted by a regional expert (Odonata Central)
% laxer vetting for iNaturalist - atleast 2 members have to agree then it is "research grade"
% assumption of occupancy modeling - sampling site indepencence - they had a lot of closely spaced sites (violates this assumption) - each site is a 1km by 1km grid (ArcMap) - resonable estiamte of max dispersal of adult odonata 
% software : ArcMap (Python), R (unmarked, rjags, segmented)
% used body length as a covariate for detection probability 
% log transformed non-categorical data - high freq low values
% piecewise regression model 

\citet{farhadinia2018citizen} & Bayesian hierarchical and  Bayesian occupancy models & Estimates of the occupancy and detection probabilities of Persian leopard species in remote landscapes   & \textsf{R} (\textsf{R2jags}), \textsf{JAGS}  &  data collected from 80 sites on 2013 on Northeastern Iran.     & Platform\\ \hline % 
% ccombines profesional monitoring and CS
% model introduced in MacKenzie et al., 2002; Tyre et al., 2003)

\citet{morii2018activity} & Bayesian and PCA regression models & Zero-inflated negative binomial model the frequencies of slug Limax maximus in Japan & \textsf{R}, \textsf{Stan} &    Data collected from one participant from 2015-2016& \\ \hline % 
% a bit weird because only one citizen 

%C	&	 &  &  \textsf{R} &   &  &\\  \hline 

\citet{bradter2018can} & Bayesian site- occupancy-detection model. Logistic regression & Assesses habitat suitability models (HSMs) of Siberian jay (Perisoreus infaustus) in Sweden & \textsf{R} (\textsf{multispeciesPP}), \textsf{JAGS} &    questionnaire answers from 60 subjects carried out from 2000-2013  & \\ \hline % 

%presence–absence
% professional data from: Swedish Species Observation System
% no spatial regression / prior

\citet{reich2018integrating} & Bayesian species occupancy models & Two-stage optimal survey design by minimization of the misclassification errors. Accounts for spatial variation. & \textsf{R} (\textsf{mgcv})   & Data from brown-headed nuthatch (Sitta pusilla) species collected in the Southeast United States & eBird\\ \hline %

% optimum experimental design. two stages : frst using CS data and then with a prof survey

\citet{petracca2018robust} & Bayesian hierarchical occupancy model & Assesses the occurrence jaguars (Panthera onca) in Central America. Accounts for spatial and temporal dependence & \textsf{R}, \textsf{jagsUI}, \textsf{JAGS} &   3863 interviews from 2009-2014 &   \\ \hline %
% used interviews
%jaguars outside protected areas in Central America
% same old model for apparent prevalence
% predictors
% big claims in this paper

\citet{coron2018bayesian} & Bayesian abundance model & Estimates the abundance of birds correcting the bias introduced in the opportunistic data  & \textsf{R} (\textsf{rjags}) &   species of bird in Aquitaine (France) &  \\ \hline %

\citet{zipkin2018synthesizing} &  Bayesian population model & Review integrative population models via data integration approach &    &  -   &  \\ \hline
% used to assess population declines

\citet{shima2018factors}      & Bayesian hierarchical model    & Assessing the impact of gender, age, weight, season, rainfall, traffic volumes, etc. on road kill risk of Lumholtz's tree kangaroo   & \textsf{R} (\textsf{R2WinBUGS})  & Citizen sightings (1998-2000) and road-kill data (2012-2017), Atherton Tablelands, Australia & Tree Kangaroo Mammal Group (data set) \\ \hline
% P-ID/YR: SHIMA, 2018
% P-FOCUS: Assessing the impact of gender, age, weight, season, rainfall, traffic volumes, etc. on road kill risk of Lumholtz's tree kangaroo.
% P--DATA: Citizen sightings (1998-2000) and road-kill data (2012-2017), Atherton Tablelands, Australia.
% P-MODEL: Bayesian hierarchical model, spatial (cite in Atherton Tablelands).
% M-EXTRA: Normal priors (diffuse). Bias mentioned, not addressed.
% P--SOFT: R (R2WinBUGS), WinBUGS
% P--PLAT: Tree Kangaroo Mammal Group (data set).

\citet{mair2017evaluating} & Bayesian species distribution model & Models the occurrence of {\it Phellinus ferrugineofuscus} fungus used to forecast national forest projections in Sweden using 
presence and/or presence–absence data. & \textsf{OpenBUGS}, \textsf{R} (\textsf{R2OpenBUGS},  \textsf{BRugs})   & 2000-2013 & Swedish Lifewatch project \url{www.analysisportal.se} \\ \hline %
% no trully spatial
%& space, time
%Bayesian occupancy model
% "`Forecasts of species occurrence from models based on citizen cience data were congruent with forecasts made using the colonization–extinction model based on systematically collected data
% 100 m grid cell resolution
% compared 5 methods. GLM,maxEnt
% they discuss variable selection.

\citet{dennis2017efficient} & Bayesian occupancy model & Compares Bayesian and frequentist occupancy models for presence-only data & \textsf{R},  \textsf{Sparta} &   UK Butterflies for the New Millennium database containing more than 11 million species occurrence. & \\ \hline

% cmpares \citet{isaac2014statistics} model
%space , time &
% data from Asher J, Fox R, Warren MS. British butterfly distributions and the 2010 target. Journal of Insect Conservation.2011; 15:291±299. https://doi.org/10.1007/s10841-010-9346-7
%UK Butterflies for the New Millennium data base

%1 km2 grid cells
% argues that non bayesian can be a better alternative. faster computation

\citet{fournier2017combining} &  Bayesian species distribution model 	&	Estimates the origins of species using stable isotopes from capture data while being informed by a species distribution model obtained from CS data.  environmental variables &	\textsf{ARCMAP}	 &  Live traps 2005 - 2009 and eBird reports (n = 3,632) from 2002 to 2012 of the Virginia rail {\it Rallus limicola}  & eBird \\ \hline
% integrates CS with capture data 
%spatial 
% bias correction
% compare benchmark stable isotopes data
% predictors: land cover, wetland presence, bioclimatic params
%1 km2 resolution
% not trully spatial 
% account for sampling bias using the samplign bias grid and targetting group sampling
% bayes rule using ARCMAP

% NOT BAYESIAN \citet{mengersen2017modelling} &   Occupancy and abundance models 	&	Estimates the spatial abundance of jaguars by using opportunistic citizen science data. They used spatial covariates and mechanistic weights to account for uncertainty in citizens' elicited data.	 &		 &  	 &   &  \\ \hline
% This paper not in websci and scopus search 

\citet{eritja2017direct} &  Bayesian logistic regression 	& Asian tiger mosquito (Aedes albopictus) alert probability (proxy for prevalence) in vehicles in Spain. 		 &	\textsf{R} (\textsf{rstanarm}), \textsf{Stan}	 &  Expert validated citizen science reports of Aedes albopictus in vehicles in Spain (2014-2016)  & Mosquito Alert mobile app, \url{http://www.mosquitoalert.com/}\\ \hline

\citet{faulkner2017using} & Extension of Dirichlet process mixture model &		 Hebridean Mink Project, which was set up with the objective of removing mink from North Uist, Benbecula and South Uist, while also trying to reduce the density from South Harris. Estimates source locations of invasions directly from spatial point pattern data without the need to specify dispersal parameters.   &	\textsf{R} (\textsf{rgeo})	 &   	  Trapping data from 2001-2005 with 409 captures. Citizen sightings from 2002-2005 with 125 sightings.  & Hebridean Mink Project \\ \hline
% Hebridean Mink Project - they want to remove mink from the environment (scottish islands)
% Extension of Dirichlet process mixture (DPM) model of geographic profiling 
% this is used to estimate source locations of invasions directly from spatial point pattern data without the need to specify dispersal parameters 
% Prior: assumed an inverse-gamma prior, set to 2, corresponds to a weakly informative distribution 
% Does not explain the maths - how do they account for spatial/temporal/bias?
% Model from a criminology model 
% model good for biological invasions in the early stages when interventions are likely to be most successful and data quality low with a poorly characterised population 
% doesn't account for bias 

\citet{granroth2017non} & Bayesian pedigree reconstruction 	& Establishing pack dynamics of grey wolf (Canis Lupus) population in south-west Finland. 			 &	\textsf{R} (\textsf{MasterBayes}, \textsf{pegas}, \textsf{adegenet}, \textsf{pedigree}) &   Local hunters and nature enthusiasts collected non-invasive samples (2013-2016). Samples processed in laboratories by experts. &  \\ \hline
% Wolves! establishing pack dynamics of grey wolf (Canis Lupus) population in south-west Finland 
% non-invasive samples (2013-2016) - supplemented genetic information from wolves that were killed or trapped (in 2008-2015) 
% non-invasive = fecal samples 
% sample collections by local wildlife agencies 
% "citizen science" - local volunteers to collect fecal samples 
% citizens mostly carnivore personnel (hunters) that have been trained in wolf tracks and signs 
% some citizens local hunters or nature enthusiasts 
% only some secretion/tissue samples (other non-invasive method) 
% DNA - can identify individual wolves 
% two error rates 1) drop out rate, 2) stochastic error 
% estimated error rates for fecal samples only 
% lack of accurate temporal information (citizens estimated age of scat) - manually corrected some pedigree assignments post hoc by biologists 
% pedigree - matching up DNA to determine parents 
% there's a cool pack area map
%  A lot of the paper is about wolves (not much discussion on the Bayesian model used) 

\citet{hill2017fine} & N-mixture models in hierarchical Bayesian framework 	& Predict population size (abundance) and spatial and temporal patterns of Bicknells's Thrush (Catherus bicknelli). 		 &	
\textsf{JAGS}, \textsf{R} (\textsf{unmarked}), \textsf{ArcGIS}	 &   14,552 five minute points counts at 747 locations by citizen scientists on hiking trails & Mountain Birdwatch (MBW) monitoring scheme \\ \hline
\citet{hof2017first} & Capture-mark-recapture (CMR), Bayesian growth models
 	& Citizen-science population abundance and growth rate estimates for green sea turtles foraging in the northern Great Barrier Reef, Australia		 &	\textsf{R} (\textsf{Rmark})	 &   Capture Mark Recapture of Green Seas Turtles in the Northern Great Barrier Reef, Australia. 2003-2014, citizen (volunteers) teams lead by experts tagged 1316 turtles.  &  \\ \hline
 	% time series analysed
%First citizen-science population abundance and growth rate estimates for green sea turtles Chelonia mydas foraging in the northern Great Barrier Reef, Australia
% tagged 1316 turtles (2003-2014) 
% capture-mark-recapture (CMR), Bayesian growth models 
% population recovering at 8.3% per year 
% citizen scientists trained in CMR techniques and overseen by at least 1 researcher or experienced collaborating partner 
% Sampling effort varied between years
% software : R (RMark)
% cool turtle property - they have high fidelity to foraging grounds so apparent survival is likely to be close to true survival 
% % Sea turtles in GBR, Australia 
% doesnt account for bias nor mentions priors/modelling - MARK (R interface) for CMR models

% cite & method & description & software & application & data & platform
%\citet{leibovici2017data} &  	& Discusses the relationship between Quality Assurance (QA) and data conflation process or data fusion (DCDF)	&		 &  Method	 &   & - \\ \hline
% about the unknown quality of crowdsourcing and social media data mining - citizen science 
% Quality Assurance (QA) and data conflation process or data fusian (DCDF) - looks at this relationship
% advocates for a decoupling of the QA process and the DCDF step as much as possible while still integrating them within an approach analogous to a Bayesian paradigm
% check citations 4, 5, 6-10
% **** discuss with Edgar - possibly remove 

\citet{mang2017accounting} & Hierarchical Bayesian approach	& Models the spatio-temporal heterogeneity in detection patterns of invasive ragweed (Ambrosia artemisiifolia)		 &		    	 &   Historical records in Austria (opportunistic data) &  \\ \hline

\citet{steger2017safari} & Hierarchical Bayesian Abundance 	& Assessing the reliability of citizen science data for mammals in the East African dry-lands.  	 &	\textsf{R}	  &  Volunteer citizen scientists from tourist lodges tagged and recorded wildlife counts that were modelled against expert sampling. & \\ \hline

\citet{arab2016spatio} & Hierarchical Bayesian spatio-temporal model  &  
It models spring arrival day in two species of migratory birds (Ruby-throated Hummingbird Archilochus colubris and Purple Martin Progne subis). It accounts for sampling effort, measurement error and a conditional Autoregressive (CAR) prior is used to model the mean & 
\textsf{OpenBUGS} &  Eastern North America from 2001 to 2010  & \url{hummingbirds.net} and \url{https://www.learner.org/jnorth/} database and Purple Martin Conservation Association \url{www.purplemartin.org}\\ \hline
% very similar to \citet{arab2015spatio}
% response variable is the arrival day
% space (CAR prior structure), time (polynomial)  & 
% covariates 'climate and weather variables such as the North Atlantic Oscillation index,winter temperature, winter–spring precipitation, sampling effort, and altitude impacted migration dates.
% temperature, precipitation, winter (December–March) values of the North Atlantic Oscillation Index (WNAO), and altitude
%"'Our predictor variables include January–February temperature (JF-Temp), February–April precipitation (FA-Precip), WNAO, mean altitude (calculated based on the mean of altitude values for each cell), and sampling effort (i.e., the number of observers in each cell). All the predictor variables were standardized.'
% no codes available
% poor spatial resolution. massive tiles

\citet{siddharthan2016crowdsourcing} & Naive Bayes classifier  &  Bayesian online updating model for consensus voting. Compares efficiency the consensus labels obtained from Naive Bayes with the majority vote algorithm &  \textsf{Weka} &  classification of images from bumblebees in the United Kingdom (22 species). 8,844 classification from 1,613 images classified and 763 users& BEEWATCH (\url{www.abdn.ac.uk/research/beewatch}) \\ \hline
% not fully Bayesian
%incremental Bayesian model 
% datsa size and quality
% consensus voting
% citizens classify images
% 
% citizens submit images and produce classifications

\citet{soykan2016population} & Bayesian log-linear hierarchical model  & Models population trends in the abundance of bird species and  investigate its association with biological traits.    &  \textsf{OpenBUGS}  \textsf{R} (\textsf{R2OpenBUGS}) &    Birds counts by more than 70,000
observers/year in North America. 551 species recorded annually from Dec 14 - Jan 5 between 1966 - 2013 &  \\ \hline
% citizens privide counts
%within a circle 24.1 km (15 miles)
% no trully spatial, no autocorrelation structure 
% 2/3 of the species found to increase the population estimates.
% no sensitivity 

\citet{armstrong2016using} &  Bayesian occupancy model 	& 	Models three turtle species occupancy at one reference site	 &	\textsf{OpenBUGS}	 & One person visiting one reference site and counting the number of basking turtles  &  \\ \hline
% occupancy models - freshwater turtles are easy to detect 
% one site was repeatedly surveyed for three turtle species 
% these models were applied to 23 other water bodies in two conservation reserves with similar habitat 
% date, time of date and environmental variables - temporal? 
% mentions:Turtle Tally and Ontario Reptile and Amphibian Atlas Project BUT doesn't use this data
% "lends" itself to citizen science - potential to be a citizen science project but isnt? 
% others have already done citizen science and turtles (discussed and referenced in paper)
% Painted Turtles (Chrysemys picta), Blanding’s Turtles (Emydoidea blandingii) and Snapping Turtles (Chelydra serpentina) in two conservation reserves
% the paper is written in first person
% randomly picked survery times between 0800 - 1800 
% counted turtle with binoculars from a vantage point 
% log-linear modeling (Poisson error term, log link function)
% ***** NOT Citizen Science - POTENTIAL for citizen science ******** 

\citet{berberich2016detection} & Nest-level data-augmentation patch-occupancy model & 
Uses event-specific covariates and a plot-level Bayesian and maximum-likelihood model.
Compared 3 different methods for quantifying detection probability of Red Wood Ant (RWA) nests (highly visible, up to 2 metres)		 & 
\textsf{JAGS}, \textsf{R} (\textsf{vegan})		 &  
60 minutes for each of 16 3600 m2 plots, classified by 8 observers  (2 expert, 6 inexperienced). GPS tracked, tagged and photos. R code available Harvard Forest Data Archive. &  \\ \hline
%specpool function
% sessile organisms - difficult to perfectly detect 
% red wood ants! (counting nests) - nests are easy to detect (up to 2 metres tall) 
% eight observers  over sixteen 3600msq plots in black forest, Germany 
% used GPS to tag nests - all observers visited each plot independently (tag and photo) to remove bias 
% photos taken to double check classification and any doubling-up 
% observers of varying experience - citizens! 
% three different statistical approaches for quantifying detection probability of sessile organisms 
% 1) nest-level data-augmentation patch-occupancy model with event-specific covariates - JAGS
% 2) a plot-level Bayesian and maximum-likelihood model - one bayesian, two maxlikelihood 
% 3) nonparametric Chao-type estimators - R (specpool function from vegan package) 
% All R code is available - Harvard Forest Data Archive 
% 147 observed nest were detected by all observers - other nests only by one single observer 
% each new observer added some additional information 
% fewer observers led to higher estimates of overlooked true nests 
% overlooked 26% of nests from plot-level Bayesian detection 
% Multiple observers detect and overlook different individual nests
% overall: Bayesian methods provide more precise estimates of detection probability
% some data augmentation - models detection probability but adds characteristic for each nest and estimates the number of unobserved nests 

\citet{dwyer2016using} & Zero-inflated Poisson with conditional auto-regressive correlation structure 	& Frequency of southern cassowary traffic strikes. Used traffic strike and citizen-collected sightings.		 &	\textsf{R} (\textsf{gstat}, \textsf{ape}, \textsf{R2WinBUGS}), \textsf{WinBUGS}	&  1km-sq grid of sightings from Community for Coastal and Cassowary Conservation, 1999-2012. Traffic strikes. &  \\ \hline

\citet{yan2016informatics} & Bayesian belief network & 
Assess ecosystem services related to water quality with Thames case study. Combines citizen science and researcher probe data.  & 
\textsf{SQL}, \textsf{R} (\textsf{gRain}) &
citizen science probe data (water quality essential parameters), researcher probe data (water quality essential parameters), citizen mobile data (waterbody features), water monitoring data from UK Environmental Agency, climate data from global model simulation (1993-2012)  &  \\ \hline
% water quality distribution in time and in space 
% Bayesian belief network to assess ecosystem services related to water quality 
% framework paper (methods?)
% case study: water quality valuation in Thames 
% data: citizen science probe data (water quality essential parameters), researcher probe data (water quality essential parameters), citizen mobile data (waterbody features), water monitoring data from UK Environmental Agency, climate data from global model simulation 
% all data collated into one database (SQL) 
% Bayesian BN 
% directed acyclic graph (DAG), conditional probability table (CPT) 
% only data in spring-summer period 
% Generalized DAG of Bayesian belief network for eutrophication in the River Thames
% the math is not discussed/explained at all - assumed you know what a Bayesian BN is 
% did not discuss any bias or priors 
% mention because this uses an unusualy approach (not regression/occupancy) more complex 
% Kerrie might like this paper 

\citet{WOS:000378787700016}      & Bayesian hierarchical model & Investigating flight patterns of western monarch butterflies over the past 40 years in California & \textsf{R} (\textsf{rjags}) &  Identification data for western monarch butterflies (1972-2014, Northern California, North America) & Art Shapiro's Butterfly Site, The Xerces Society, North American Butterfly Association, Monarch Net \\ \hline
% P-ID/YR: ESPESET, 2016
% P-FOCUS: Investigating flight patterns of western monarch butterflies over the past 40 years in California.
% P--DATA: Identification data for western monarch butterflies (1972-2014, Northern California, North America).
% P-MODEL: Bayesian hierarchical model, spatio-temporal (sites in Northern California per year from 1972-2014).
% M-EXTRA: Unspecified priors (uninformative). No mention of bias.
% P--SOFT: R (rjags), R-JAGS.
% P--PLAT: Art Shapiro's Butterfly Site, The Xerces Society, North American Butterfly Association, Monarch Net.

\citet{purse2015landscape} & Bayesian spatial survival model &  Time to invasion from species as a function of 
environmental predictors & \textsf{INLA} \textsf{R} (\textsf{survival}) &  ladybird Harmonia axyridis in Great Britain & UK Ladybird Survey (UKLS) \url{http://www.harlequin-survey.org/}  \url{http://www.ladybird-survey.org/}  \\ \hline %
%spatio temporal & 
% the accelerated failure time (AFT) 
%"survival models over ordinary least squares regression for modelling timing of invasion is that data can
% be included from areas that have not been invaded by the end of the study period – so-called censored data"'

% experts identified the species from field visits, from specimens and photo submitted by citizens 
%10-km grid squares

% spatial random effect for the invasion probability is added 

\citet{barnes2015evaluating} & Bayesian logistic regression & Populations modelling in 21 species
of birds for assessing the efficacy of protected areas  &  \textsf{R} (\textsf{R2JAGS}, \textsf{LISZT}), \textsf{JAGS},  & Wet Tropics in Queensland, Australia & New Atlas of Australian Birds \\ \hline %
%predictor indicator: protected area status
%time
%Sixteen species were more likely to be found within protected areas, two more likely to be recorded outside, and three
% showed no statistically significant difference

\citet{arab2015spatio} & Hierarchical Bayesian spatio-temporal model  & Spring arrival dates in bird migration from Purple Martins from (1905-1940) and (2001-2010). Error processes modelled with an auto-regressive model AR(1) and a Conditional Autoregressive (CAR) prior was used for the spatial dependency &
\textsf{OpenBUGS}	&  initial data was collected by competent
naturalists from the North American Bird Phenology Program (NABPP) and more recently by the Purple Martin Conservation Association (\url{www.purplemartin.org}).    & \\ \hline
% response var is arrival day of the year. 
% space (CAR prior structure), time (AR(1)) &
% no sensitivity analysis
%North American Bird Phenology Program and Purple Martin Conservation Association
% limitations: grid size is enormous limiting the power of the spatial CAR prior.
% codes BUGS are given

\citet{yoshikawa2015dietary} & Bayesian hierarchical time series model   & Models the month proportion of dietary items using long term monitoring data collected by citizens. 
The authors found the results of the Bayesian approach to be more consistent with other studies available literature.
&  \textsf{R}, \textsf{JAGS} &    15  bird species in Kanagawa, Japan & \\ \hline
% citizens reported if the bird were eating or not aND  the type of food (10 categories). 
%temporal autocorrelation AR(1) &
% from 1977 - 2006  
% no spatial
%It combines multiple approaches with direct observation

% non Bayesian \citet{mcclintock2015endangered} &  &  abundance Florida panthers (\emph{Puma concolor coryi}) populations & &   & motor vehicle collisions reported in Florida, USA  & \\ \hline

\citet{chapman2015unbiased} & Naive Bayes with discrete Fourier transform 	& Opportunistic data from two botanical gardens for flowering periods.		 &	 \textsf{R} (\textsf{MCMCpack}) 	 & Botanical Society of Britain and Ireland (BSBI)   &  National Biodiversity Network (NBN) \url{https://data.nbn.org.uk/}\\ \hline
% Biological recording - ***opportunistic citizen science*** in which observations are collected without standardized sampling protocols 
% naive model that ignores recording effort and assumes seasonal pattern in recording of plant phenology
% both models are based on the discrete Fourier transform (DFT) - annual cycles
% uses Fourier terms to account for periodical changes in the time series (harmonic covariates) 
% Botanical Society of Britain and Ireland (BSBI), National Biodiversity Nework (NBN) https://data.nbn.org.uk/
% temporal - date is in Julians 
% likelihood function of recording-corrected model 
% flat improper priors for all parameters, metropolis sampling of MCMC 
% Robust Major Axis (MA) regression to test relationships in obs and estimates 
% 20 species from RBG Edinburgh, 9 from RBG Kew (data) then removed 4 species because "clear outliers" 
% worried about recording bias - from raw recording dates 

\citet{pagel2014quantifying} & Bayesian hierarchical spatio temporal approach for modelling the abundance of species    & Data integration from profesional and CS programs to achienve improved estimates.  & \textsf{OpenBUGS} & butterfly species (Pyronia tithonus) from Great Britain from 1985 to 2004. &  New Millennium (BNM) project \url{https://butterfly-conservation.org/our-work/recording-and-monitoring/butterflies-for-the-new-millennium}\\ \hline %
%space and time & 
%spatially explicit state-space model of abundance 
% we did not include any environmental covariates. jsut spatial randomeffects

\citet{lottig2014long} &  Bayesian regression &  Assesses trends in lake-water clarity using a Bayesian hierarchical modeling. & \textsf{JAGS} &    239,741 Secchi depth measurements by citizens in the United States from 1938–2012 & \url{https://www.nalms.org/secchidipin/} \\ \hline %
%temporal change &
% in water quality monitoring 
% not sure if ecology
% temporal change. the year is a covariate
% predictors: predictor variables (i.e., latitude, lake surface area, median sample date)

\citet{strebel2014studying} &  Bayesian occupancy model & Penalized thin-plate splines accounting for space and time & \textsf{Jags} &   approx 1800 skilled citizens from the Swiss citizen-science bird recording scheme IS& \\ \hline %

% presence - absence
% it does not account for false positive detections.
% use meta analysis to combine info form species

\citet{roth2014estimating} & Bayesian site-occupancy model 	&	Estimates the average arrival and departure dates from a single breeding season for two species of long-distance migrant songbirds and two species of butterflies	 &	 \textsf{OpenBUGS}, \textsf{R}	  &  Simulation &  \\ \hline
% uses simulated data
% spring arrival of two long-distance migrant songbirds
% length of flight period of two butterfly species, marbled white, the ringlet
% Switzerland long term monitoring program  
% method 'could' be applied to temporally and spatially extensive data for citizen science 
% for birds: arrival to breeding grounds, Garden Warbler, Spotted Flycatcher
% two new parameters to site-occupancy model. 1) probability that the species enters the study area between sampling occasions 2) probability that the species will depart from the sites before the next sampling occasion 
% estimates average arrival and departure dates from a single breeding season  - assumes population closure upon arrival 
% true occupancy state is a Bernoulli random variable
% arrival and departure times of species is an over-dispersed Poisson distribution 
% **** Not citizen science - potential to be citizen science ****

\citet{isaac2014statistics} & Bayesian occupancy model & Compare 11 occupancy methods in terms of the efficiency to detect trends based on different opportunistic data collection scenarios  &  \textsf{JAGS}  &  Simulated data with parameters obtained from previous studies.  & \\ \hline %
% really good paper (200 citations). re read
% not spatial random effect
% model in Occupancy_model.r    C:\Users\santosfe\OneDrive - Queensland University of Technology\projects\SC\papers CS Bayes\Isaac RangeChangeSims-master\RangeChangeSims-master\Occupancy\

\citet{beck2014using} & Jolly–Seber mark-recapture model  & Implemented a population dynamics model considering recruitment and losses from individuals identified in images  &  \textsf{WinBUGS} &  opportunistic encounters reported by citizens and whale watching communities  in UK and Irish waters between 1992 and 2011  & \\ \hline
% citizens contribute the images
%Bayesian formulation of the model (Royle & Dorazio, 2008; Fearnbach et al., 2012)
% temporal component is not explisitly included
%'by the Hebridean Whale and Dolphin Trust and the Irish Whale and Dolphin Group'
%"'Bayesian formulation of the Jolly–Seber mark-recapture model to the photo-identification data compiled from opportunistic photographic encounters with this population of killer whales"'
% opportunistic
% no bias correction.
% no sensitivity analysis
% no covariates

\citet{van2013opportunistic} & Bayesian hierarchical occupancy model & Estimation of the annual occupancy probability. Accounts for changes in space and time and corrects geographical and observation
effort biases.  & \textsf{R} (\textsf{Rjags})  &   
Butterfly (1990–2010) and dragonfly (1999–2011) species in the Netherlands obtained from databases such as National Database Flora and Fauna, Dutch Butterfly Monitoring Scheme, and Dutch Dragonfly Monitoring Scheme. 
&  \\ \hline %

% Adpated the model by \citet{royle2008hierarchical}.
% occucpancy probability is cumputed as a function of the occupancy at time $t-1$ and the persistence (species will remain in the site) and colonization probability 9species will colonize the site after being absent).
% they reported similar distribution trends using the oportunistic derived data compared to the professional programs.

% not really spatial. 

\citet{cheney2013integrating} & Bayesian distribution model & Hierarchical log-linear approach for modelling bottlenose dolphins ({\it Tursiops truncatus}) in Scotland using data integration from multiple sources &  \textsf{WinBUGS}  & species identified from images from general population and wildlife tours (2006 - 2007) along with multiple professional data sources   & \\ \hline %
% & space. 
% data integration
%Bayesian multi-site mark-recapture model to estimate abundance distribution, movements and abundance of
% photo identification
 % review
% no codes provided

\citet{wilson2013citizen} &  Bayesian hierarchical model. Log linear model 	&	It models the abundance of Western Grebes in the Salish Sea.
Species counts are described using a Poisson distribution and includes covariates to account for the sampling effort and and population effects.   &	\textsf{WinBUGS}	  &  Audubon Christmas Bird Count data	(36 years) with almost 2.5 million records  &  \\ \hline
% count data
%&  region specific intercepts and year specific slopes
% 24.1 km count circle
%mid-December to early January
%accounts for space and time as a random effect.
% very simple model
% assess the decline in abundace of species 
% Spatial and temporal variation is considered by means of  region specific intercepts and year specific slopes    

\citet{penone2013use} & Bayesian model averaging (BMA) for general linearized models 	&	Detect widely known patterns of anthropogenic effects on Bush-Crickets. 	 &	
\textsf{R} (\textsf{spdep}), \textsf{ArcGIS}, \textsf{Syrinx}	 & 
 Citizen collect data, expert classified data.  	 & 
French Museum of Natural History  \url{http://vigienature.mnhn.fr} \\ \hline

\citet{van2013occupancy} & Bayesian occupancy model with hierarchically coupled submodels for occupancy and detection 	& 
Damselfy (Calioteryx splendens) opportunistic data from 5 countries  to estimate the annual number of occupied sites per country. Countries are weighted by number of sites surveyed and range of species per country, this adjusts for unequal geographical distributions.		 & 
\textsf{JAGS}, \textsf{WinBUGS}		 &  
Opportunistic data - from many sources, no standardised sampling design. 1990-2008, Ireland, Great Britain, Netherlands, Belgium, France.   & Many \\ \hline
% damselfly Calioteryx splendens 
% occupancy modelling 
% data from Ireland, Great Britain, Netherlands, Belgium, France - 5 countries 1990-2008
% occupancy models to estimate the annual number of occupied sites per country
% each site is 1km x 1km 
% presence-absence data - occupancy - account for imperfect detection 
% model per country in JAGS 
% weighted countries by number of sites surveyed and range of species per country - adjusts for unequal geographical distribution of sites 
% opportunistic data - from many sources, no sample design, single records for a particular species 
% presence-absence - 1 for damselfly, 0 for any other species - restricted for Julian dates 130-250
% occupancy models consists of two hierarchically coupled submodels - one for occupancy, one for detection (conditional on occupancy)
% Many sources of data (atleast 1 per country) 
% The country weights are the 'interesting' part of this paper 
% van Strien released two papers in 2013 on the similar data/model - compare them 

\citet{meiman2012comparing}      & Bayesian hierarchical model & Prediction of saltmarsh sparrow distribution and nesting activity using multiple Bayesian hierarchical models & \textsf{WinBUGS} &  Marsh classification data (2006-2008), saltmarsh sparrow presence data (2006-2008), Connecticut, North America &  \\ \hline
% P-ID/YR: MEIMAN, 2012
% P-FOCUS: Prediction of saltmarsh sparrow distribution and nesting activity using multiple Bayesian hierarchical models.
% P--DATA: Marsh classification data (2006-2008), saltmarsh sparrow presence data (2006-2008), Connecticut, North America.
% P-MODEL: Bayesian hierarchical model, spatial (site in North America).
% M-EXTRA: Normals priors (vague). No mention of biases.
% P--SOFT: WinBUGS.
% P--PLAT: N/A.

\citet{snall2011evaluating} & Bayesian hierarchical regression model  & Poisson log-linear model for modelling the species abundance with autorregressive component for temporal dependency . Compares abundance trends obtained from a professional and citizen science dataset using presence-only data. A mechanistic weighting is used to account for survey effort &  \textsf{OpenBUGS} &  Species Gateway and the Swedish Bird Survey from 2001-2009 in Sweden & \url{http://www.artportalen.se}\\ \hline

% 242 bird species from 2001-2009 
% citizens performed the survey. 
% Explanatory variables
% & autorregressive model for temporal variations
% reg coefficients were estimated for every species group (forrest, marine, etc)
% found negative assoscation between species presence using profesional and CS data. highlight the improtance for verification of data and bias correction.
% no spatial

% model details: Snall 1-s2.0-S0006320710004805-mmc1
% no sensitivity analysis

\citet{szabo2010regional} & Bayesian logistic regression model for List Length Analysis & Models the presence/absence of 269 species as a function of the year and the length of the list &  \textsf{WinBUGS}  & bird lists collected by Birds Queensland over 40 years around Brisbane, Australia   & \\ \hline

% no spatial or temporal dependency
%& time is a predictor 

\citet{kadoya2010predicting}      & Bayesian ecological-niche model    & Using spatio-temporal presence/absence data to predict control of the near-future spread of the bumblebee   & \textsf{R}, \textsf{ArcGIS}  &  Precence/absence data for B. terrestris (1992-2010, Hokkaido, Japan) & B. terrestris monitoring program \\ \hline
% P-ID/YR: KADOYA, 2010
% P-FOCUS: Using spatio-temporal presence/absence data to predict control of the near-future spread of the bumblebee.
% P--DATA: Precence/absence data for B. terrestris (1992-2010, Hokkaido, Japan).
% P-MODEL: Bayesian ecological-niche model, spatio-temporal (sites in Hokkaido per year from 1992-2010).
% M-EXTRA: Uniform and Normal priors (vague). No mention of biases.
% P--SOFT: R, ArcGIS
% P--PLAT: B. terrestris monitoring program.

%\citet{WOS:000632445400001}      & METHOD    & DESCRIPTION   & SOFTWARE/PACKAGE  & APPLICATION   & DATA    & PLATFORM/PROJECT \\ \hline
% P-ID/YR: BURIAN, 2021 (not included: study only includes simulated data)
% P-FOCUS: Assessing rates of false positive and negative results in spatial data-structures by reviewing advances in occupancy models.
% P--DATA: 
% P-MODEL: Bayesian occupancy model
% M-EXTRA:
% P--SOFT:
% P--PLAT:

\label{table:summ}
\end{longtable}

\end{landscape}

\clearpage
\restoregeometry

\end{document}